\documentclass{emulateapj} 
\usepackage{times}

\newcommand{\D}[2]{\frac{\partial #2}{\partial #1}}
\newcommand\bb[1]{\mbox{\boldmath{$#1$}}}
\newcommand\del{\bb{\nabla}} 
\newcommand\bcdot{\bb{\cdot}}
\newcommand\btimes{\bb{\times}}

\begin{document}

\shorttitle{\textsc SATURATION OF MHD TURBULENCE IN ACCRETION DISKS}
\shortauthors{\textsc PESSAH}

\title{Angular Momentum Transport in Protoplanetary and Black-Hole
  Accretion Disks: \\ The Role of Parasitic Modes in the Saturation of
  MHD Turbulence}

\author{Martin E.  Pessah} \affil{Institute for Advanced Study,
  Princeton, NJ, 08540, USA; mpessah@ias.edu}

\begin{abstract}
  The magnetorotational instability (MRI) is considered a key process
  for driving efficient angular momentum transport in astrophysical
  disks.  Understanding its non-linear saturation constitutes a
  fundamental problem in modern accretion disk theory.  The large
  dynamical range in physical conditions in accretion disks makes it
  challenging to address this problem only with numerical simulations.
  We analyze the concept that (secondary) parasitic instabilities are
  responsible for the saturation of the MRI.  Our approach enables us
  to explore dissipative regimes that are relevant to astrophysical
  and laboratory conditions that lie beyond the regime accessible to
  current numerical simulations. We calculate the spectrum and
  physical structure of parasitic modes that feed off the fastest,
  exact (primary) MRI mode when its amplitude is such that the fastest
  parasitic mode grows as fast as the MRI.  We argue that this
  ``saturation'' amplitude provides an estimate of the magnetic field
  that can be generated by the MRI before the secondary instabilities
  suppress its growth significantly.  Recent works suggest that the
  saturation amplitude of the MRI depends mainly on the magnetic
  Prandtl number.  Our results suggest that, as long as viscous
  effects do not dominate the fluid dynamics, the saturation level of
  the MRI depends only on the Elsasser number $\Lambda_\eta$.  We
  calculate the ratio between the stress and the magnetic energy
  density, $\alpha_{\rm sat}\beta_{\rm sat}$, associated with the
  primary MRI mode.  We find that for $\Lambda_\eta >1$
  Kelvin-Helmholtz modes are responsible for saturation and
  $\alpha_{\rm sat}\beta_{\rm sat} = 0.4$, while for $\Lambda_\eta <
  1$ tearing modes prevail and $\alpha_{\rm sat}\beta_{\rm sat} \simeq
  0.5 \, \Lambda_\eta$.  Several features of numerical simulations
  designed to address the saturation of the MRI in accretion disks
  surrounding young stars and compact objects can be interpreted in
  terms of our findings.
\end{abstract}


\section{Introduction}
\label{sec:intro}

The transport of mass and angular momentum in accretion disks remains
one of the least understood processes in modern astrophysics.  The
standard accretion disk model \citep{SS73, LBP74, FKR02} is based on
the assumption that turbulence provides an efficient mechanism for
enabling accretion but magnetic fields, thought to be crucial for
driving the turbulence, do not play an explicit role.  It is currently
believed that the magnetorotational instability (MRI;
\citealt{Velikhov59, Chandrasekhar60, BH91,BH98}) is responsible for
driving the magnetohydrodynamic (MHD) turbulence required for
efficient angular momentum transport in astrophysical disks.  However,
at present, there are no accretion disk models that incorporate the
physics driving angular momentum transport in a self-consistent way.

Since the appreciation of the relevance of the MRI to accretion disks,
significant progress has been made in understanding the physics of the
instability and characterizing the ensuing turbulent state. A large
set of numerical simulations \citep[see, e.g.,][]{HGB95,
  Brandenburgetal95, MS00, FSH00, SI01, SIM98, Sanoetal04, TSD07,
  FPLH07, LL07, Obergaulingeretal09} have provided insight into the
turbulent MHD flows under various physical conditions.  However, the
fundamental processes that determine the strength of the turbulence in
the non-linear regime are yet to be deciphered. At present there have
only been a handful of studies addressing the theoretical aspects of
this problem \citep[see, e.g.,][]{GX94, KJ05, UMR07, URM07, TD08, JJK08a,
  JJK08b, LLB09, Vishniac09, PG09}. Thus, understanding the mechanisms
that lead to the saturation of the MRI constitutes a fundamental
problem in modern accretion physics.

Building models that capture the physics of the MRI and its saturation
is crucial for constructing angular momentum transport models, and
global disk models, beyond the standard viscous accretion disk
\citep[see, e.g.,][]{KY95,Ogilvie03,PCP06b}.  The identification, and
eventual understanding, of correlations and scaling laws born out of
the synergy between numerical simulations, analytical, and
semi-analytical work \citep{PCP06a, PCP07, PCP08, BPV08, HB08, LBL09}
provides important insight toward this goal. This could also provide a
fruitful approach toward building sub-grid models for related
microphysical processes which are unfeasible to incorporate in global
simulations in a direct way \citep[see, e.g.,][]{AMY09}.

\citet{PG09} provided a summary of a parametric study of
incompressible, MRI-parasitic instabilities in dissipative regimes
accessible to current numerical simulations.  They stated that the
fastest growing modes are related to Kelvin-Helmholtz and tearing mode
instabilities.  In this work, we provide the details of the solutions
to the differential equations involved and solve for the fastest
growing modes in order to provide support for these assertions.  We
identify the existence of a critical Elsasser number of order unity
and show that Kelvin-Helmholtz and tearing modes dominate in the
quasi-ideal and resistive MHD regimes, respectively.  By means of a
systematic study of the parameter space involved, we reveal scaling
laws that describe the behavior of the fastest growing parasites in
these limits.  The joint analysis of the asymptotic behavior of the
MRI and the parasitic modes provides insight into the characteristics
of a viable saturation mechanism in regimes that are relevant to
astrophysical \citep[see, e.g.][]{Jin96, Gammie96, SM99, SW05, BS05,
  BH08} as well as experimental environments \citep[see,
e.g.][]{JGK01, GJ02, Sisanetal04, LGJ06, SSW03}.


\begin{figure}
\begin{center}
  \includegraphics[width=0.75\columnwidth,trim=0 0 0 0]{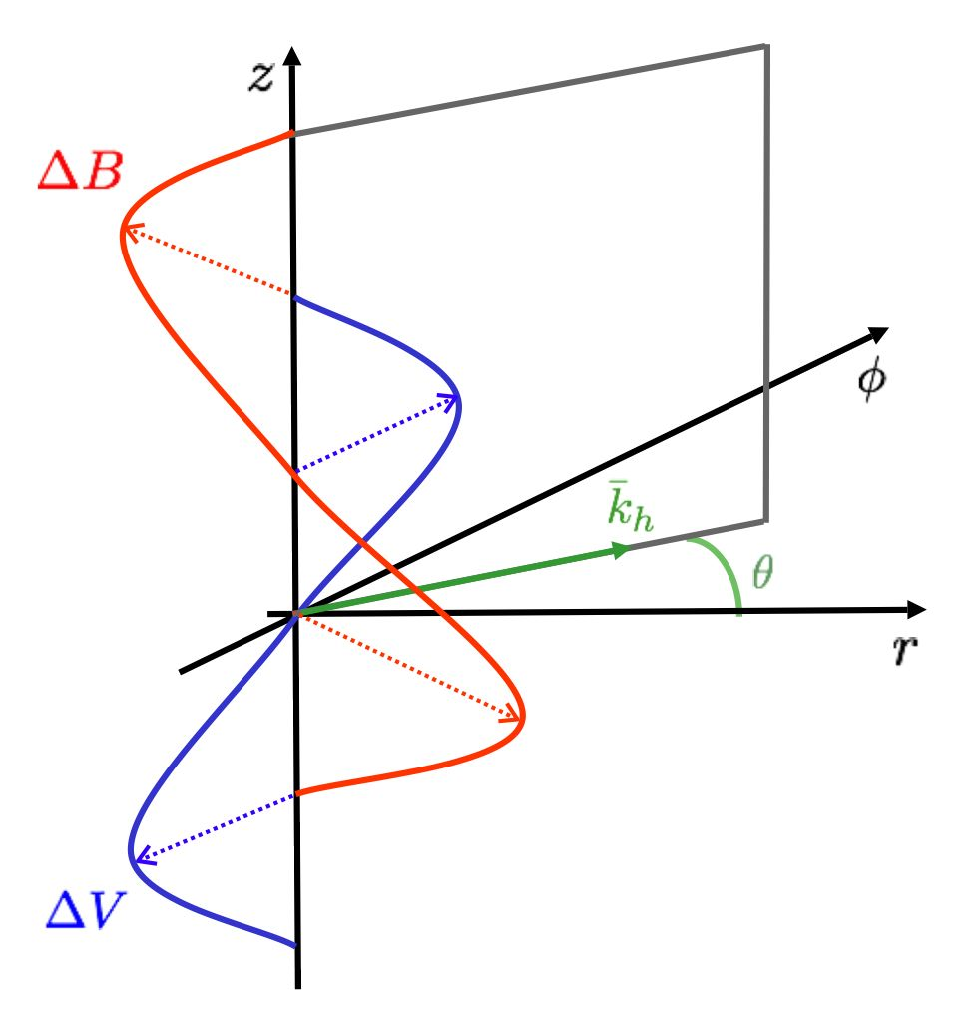}
  \caption{Three-dimensional representation of a primary MRI mode. The
    perturbations over the background Keplerian velocity and magnetic
    fields are given by $\Delta v = V_0 (\cos\theta_{\rm V},
    \sin\theta_{\rm V}, 0) \sin(Kz)$ and $\Delta B = B_0
    (\cos\theta_{\rm B}, \sin\theta_{\rm B}, 0) \cos(Kz)$,
    respectively. For a given primary MRI mode with wavenumber $K$, 
    the orientation of the planes containing $\Delta v$ and $\Delta B$ 
    and the ratio of the amplitudes $V_0/B_0$ are functions of the 
    viscosity and resistivity. The wavevector
    $\bb{k}_{\rm h}$ characterizes the horizontal wavelength of a
    given parasitic mode and $\theta$ denotes the angle between this
    vector and the radial direction, see also
    Fig.~\ref{fig:mri_representation_2d}.}
  \label{fig:mri_representation_3d}
\end{center}
\end{figure}

\section{General Considerations and Primary MRI Modes}
\label{sec:assumptions}

Consider a cylindrical background characterized by an angular velocity
profile $\bb{\Omega}=\Omega(r)\check{\bb{z}}$, threaded by a vertical
magnetic field $\bb{\bar{B}} = \bar{B}_z \check{\bb{z}}$.  We work in
the incompressible limit which is relevant when the magnetic fields
involved are so weak that the saturation of the MRI occurs at magnetic
energies that are small compared to the thermal energy. We consider
non-ideal effects due to a kinematic viscosity $\nu$ and resistivity
$\eta$, both of which are assumed to be constant.  The equations
governing the local dynamics of this MHD fluid in the shearing box
approximation are given by
\begin{eqnarray}
\label{eq:euler}
\D{t}{\bb{v}} + \left(\bb{v}\bcdot\del\right)\bb{v} & = &
- 2 \bb{\Omega}_0 \btimes \bb{v} \, + 
\, q \Omega^2_0\del(r-r_0)^2 \nonumber \\
&-& \frac{1}{\rho}\del\left(P + \frac{\bb{B}^2}{8\pi}\right)  +
\frac{(\bb{B}\bcdot\del)\bb{B}}{4\pi\rho} + \nu \del^2{\bb{v}} \,,
\nonumber \\ \\
\label{eq:induction}
\D{t}{\bb{B}} + \left( \bb{v} \bcdot \del \right)\bb{B} 
& = & \left(\bb{B} \bcdot \del \right) \bb{v} + \eta \del^2{\bb{B}} \,, 
\end{eqnarray}
where $P$ is the pressure, $\rho$ is the density, and the factor
\begin{eqnarray}
q\equiv-\left.\frac{d\ln\Omega}{d\ln r}\right|_{r_0} \,,
\end{eqnarray}
parametrizes the magnitude of the local shear at the fiducial radius
$r_0$. The continuity equation reduces to $\del \bcdot \bb{v}=0$ and
there is no need for an equation of state since the pressure can be
determined from this condition.


\begin{figure}
\includegraphics[width=\columnwidth,trim=0 0 0 0]{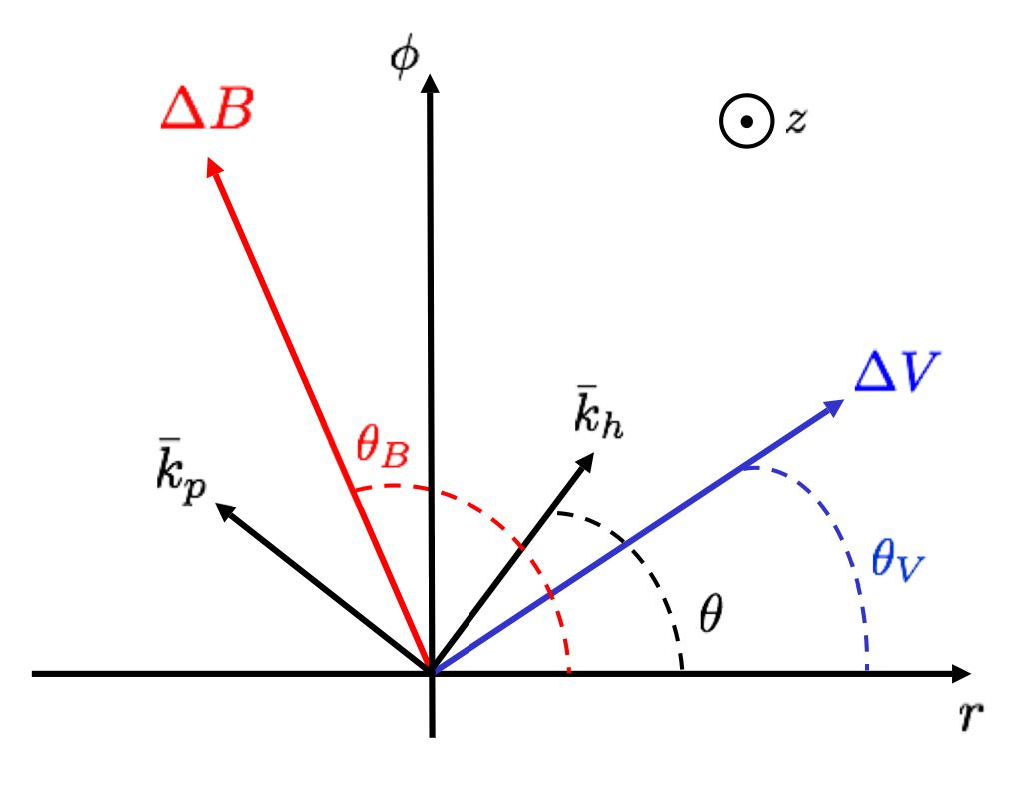}
  \caption{Two-dimensional representation of the projection of a
    primary MRI mode onto the $(\check{r}, \check{\phi})$ plane. The
    vectors $\bb{V}_0 = V_0 (\cos\theta_{\rm V}, \sin\theta_{\rm V},
    0)$ and $\bb{B}_0 = B_0 (\cos\theta_{\rm B}, \sin\theta_{\rm B},
    0)$ represent the projection of the MRI velocity and magnetic
    fields, with associated direction angles $\theta_{\rm V}$ and
    $\theta_{\rm B}$.  The versors $\check{\bb{k}}_{\rm h}$ and
    $\check{\bb{k}}_{\rm p}$ characterize, respectively, the
    directions parallel and perpendicular to the horizontal wavevector
    $\bb{k}_{\rm h}$ of a given parasitic mode.}
  \label{fig:mri_representation_2d}
\end{figure}

The set of Equations~(\ref{eq:euler}) and (\ref{eq:induction})
has solutions of the form
\begin{eqnarray}
\bb{v}&=&- q \Omega_0 (r-r_0) \check{\bb{\phi}} +  \bb{V}_0 \sin(Kz) \, e^{\Gamma t} \,,\\ 
\bb{B}&=& \bar{B}_z \check{\bb{z}} +  \bb{B}_0 \cos(Kz) \, e^{\Gamma t} \,.
\end{eqnarray}
For a given wavenumber $K$ perpendicular to the disk mid-plane, the growth
rate $\Gamma$ satisfies the dispersion relation
\begin{equation}
(K^2\bar{v}_{{\rm A}z}^2 + \Gamma_\nu \Gamma_\eta)^2 
+ \kappa^2 (K^2\bar{v}_{{\rm A}z}^2 + \Gamma_\eta^2) 
- 4 K^2\bar{v}_{{\rm A}z}^2\Omega_0^2 = 0 \,,
\end{equation}
where $\Gamma_\nu \equiv \Gamma + \nu K^2$, $\Gamma_\eta \equiv \Gamma
+ \eta K^2$, $\kappa\equiv\sqrt{2(2-q)}\Omega_0$ is the epicyclic
frequency, $\bar{v}_{{\rm A}z}\equiv\bar{B}_z/\sqrt{4\pi\rho}$ is the
Alfv\'en speed, and $\Omega_0$ is the local Keplerian frequency.  The
ratio of the amplitudes $V_0/B_0$ of the vectors characterizing the
MRI velocity and magnetic fields,
\begin{eqnarray}
\label{eq:V0_MRI}
\bb{V}_0 &\equiv& V_0\,(\cos\theta_{{\rm V}}, \sin\theta_{{\rm V}}, 0)  \,, \\
\label{eq:B0_MRI}
\bb{B}_0 &\equiv& B_0\,(\cos\theta_{{\rm B}}, \sin\theta_{{\rm B}}, 0) \,,
\end{eqnarray}
as well as their directions $\theta_{{\rm V}}$ and $\theta_{{\rm B}}$,
see Figures~\ref{fig:mri_representation_3d} and
\ref{fig:mri_representation_2d}, are known functions of
$(\nu,\eta,K)$, see \citet{PC08} for a detailed discussion. The growth
rate $\Gamma$ has a unique maximum, $\Gamma_{\rm max}(\nu,\eta)$, at
$K=K_{\rm max}(\nu,\eta)$.

Unless otherwise mentioned, we work with dimensionless variables
defined in terms of the characteristic length and time scales set by
the background Alfv\'en speed and the local angular frequency: $L_0
\equiv \bar{v}_{{\rm A}z}/\Omega_0$ and $T_0 \equiv 1/\Omega_0$.  We
subsume the effects related to viscosity and resistivity into the
dimensionless quantities
\begin{eqnarray}
  \Lambda_\nu  &\equiv& \frac{\bar{v}_{{\rm A}z}^2}{\nu\Omega_0} \,, \\
  \Lambda_\eta &\equiv& \frac{\bar{v}_{{\rm A}z}^2}{\eta\Omega_0} \,, 
\end{eqnarray}
whose ratio is the magnetic Prandtl number, ${\rm Pm} \equiv \nu/\eta
\equiv \Lambda_\eta/\Lambda_\nu$.  The quantity $\Lambda_\eta$ is
known as the Elsasser number, while its viscous counterpart
$\Lambda_\nu$ is related to the Reynolds number.  In our dimensionless
variables, magnetic field strengths are defined relative to the
background field $\bar{B}_z$, while $\Lambda_\nu \equiv \nu^{-1}$ and
$\Lambda_\eta \equiv \eta^{-1}$.

\section{Stability of MRI Modes}
\label{sec:parasitic instabilities}

The exact equations for the dynamical evolution of the velocity and
magnetic fields corresponding to secondary instabilities $\delta
\bb{v}(\bb{x},t)$ and $\delta \bb{B}(\bb{x},t)$ affecting an MRI mode
are obtained by substituting in Equations~(\ref{eq:euler}) and
(\ref{eq:induction}) the ansatz
\begin{eqnarray}
\bb{v}&=&- q \Omega_0 (r-r_0) \check{\bb{\phi}} +  \bb{V}_0 \sin(Kz) \,e^{\Gamma t} + \delta \bb{v}\,,\\ 
\bb{B}&=& \bar{B}_z \check{\bb{z}} +  \bb{B}_0 \cos(Kz) \, e^{\Gamma t}+ \delta \bb{B} \,.
\end{eqnarray}
This procedure leads to a set of partial differential equations in
space and time whose solution is beyond the scope of the present
paper. It is possible, however, to gain insight into the growth rates
and physical properties of the secondary instabilities by adopting
some simplifying assumptions.

\subsection{Assumptions}

In order to make the problem of the stability of the non-ideal MRI
modes against parasitic instabilities more tractable we adopt the same
set of assumptions and approximations stated in \citet{PG09}, see also
\citet{LLB09}, which are the same as those adopted by \citet{GX94} for
ideal magnetohydrodynamics (MHD).  The central simplification reduces
to assuming that the exact (primary) MRI modes can be considered as a
time-independent background from which the (secondary) parasitic modes
feed off.  Because the growth rates of the parasitic modes increase as
the amplitude of the primary mode increases, this approximation is
better satisfied when the amplitude of the MRI modes is large compared
to the background vertical field.  In this case, we can also neglect
the influence of the weak vertical background field, the Coriolis
force, and the background shear flow on the dynamics of the secondary
modes.

There is a subtlety associated with this assumption that is worth
stating explicitly. The amplitude of the MRI cannot be assumed to be
arbitrarily large. If this were the case, the parasites would drain an
amount of energy comparable to the energy of the primary mode
exceedingly fast. This would call into question whether the primary
mode would have been able to reach the assumed amplitude in the first
place.  There exists a regime in which the amplitude of the primary
modes is large enough that they can be taken as time-independent, but
no so large that the fast-growing secondaries would have prevented the
MRI from reaching the assumed amplitude.

Here, we are interested in estimating the amplitude to which the MRI
can grow to in order for the fastest parasitic modes to have growth
rates that are comparable to that of the primary mode upon which they
feed.  The motivation to calculate this ``saturation'' amplitude, i.e.,
$B_0^{\rm sat}$, is that the parasites will be able to drain an amount
of energy of order $(B_0^{\rm sat})^2$ from the primary modes shortly
after their growth rates are comparable to that of the primary modes.
We stress that the value of $B_0$ at which the amplitude of the
parasites is similar to that of the primary MRI mode upon which they
feed, is estimated to be a factor of a few larger that $B_0^{\rm
  sat}$, see the discussion in \citet{PG09}. Thus, it does not seem
consistent to assume ab initio that the MRI modes can reach amplitudes
$B_0 \gg B_0^{\rm sat}$.  The price that we pay for working in the
regime $B_0 \simeq B_0^{\rm sat}$ is that the assumption of a
stationary background is only marginally satisfied.

\subsection{Equations of Motion for the Parasites}

Under the assumptions stated above, the equations governing the
dynamics of the secondary instabilities are
\begin{eqnarray}
\label{eq:euler_pi}
\partial_t\delta\bb{v} &+& \left(\Delta\bb{v}\bcdot\del\right)\delta\bb{v} +
\left(\delta\bb{v}\bcdot\del\right)\Delta\bb{v}  = 
- \del(\delta P + \Delta\bb{B} \bcdot \delta\bb{B})
\nonumber \\ 
&+& (\Delta\bb{B}\bcdot\del)\delta\bb{B} 
+ (\delta\bb{B}\bcdot\del)\Delta\bb{B} 
+ \nu \del^2{\delta \bb{v}} \,,
\\
\label{eq:induction_pi}
\partial_t \delta\bb{B} &+& 
\left(\Delta\bb{v} \bcdot \del \right)\delta\bb{B} +
\left(\delta \bb{v} \bcdot \del \right)\Delta\bb{B} =
\nonumber \\ 
&& \left(\Delta\bb{B} \bcdot \del \right) \delta\bb{v} +
\left(\delta\bb{B} \bcdot \del \right) \Delta \bb{v} + \eta \del^2{\delta\bb{B}} \,,
\end{eqnarray}
where $\del \bcdot \delta\bb{v} = \del \bcdot \delta\bb{B} =0$,
$\delta P$ stands for the pressure perturbation, and we have defined
the time-independent amplitude of an unstable MRI mode with wavenumber
$K$ as
\begin{eqnarray}
\label{eq:Delta_V_MRI}
  \Delta{\bb{v}} &\equiv& \bb{V}_0 \sin(Kz) \,,  \\
\label{eq:Delta_B_MRI}
  \Delta{\bb{B}} &\equiv& \bb{B}_0 \cos(Kz) \,.
\end{eqnarray}

We seek solutions to Equations~(\ref{eq:euler_pi}) and
(\ref{eq:induction_pi}) of the form
\begin{eqnarray}
\label{eq:v_sol}
\delta \bb{v}(\bb{x},t) &=& \delta{\bb{v}}_0(z) \, \exp{[s t - i\bb{k}\bcdot\bb{x}]} \,, \\ 
\label{eq:b_sol}
\delta \bb{B}(\bb{x},t) &=& \delta{\bb{B}}_0(z) \, \exp{[s t - i\bb{k}\bcdot\bb{x}]} \,,
\end{eqnarray}
Here, $\bb{k} = \bb{k}_{\rm h} + k_z \check{\bb{z}}$, where
$\bb{k}_{\rm h} \equiv k_x \check{x} + k_y \check{y} \equiv k_{\rm
  h}(\cos\theta \, \check{x} + \sin\theta \, \check{y})$ is a
horizontal wavevector, see Figures~\ref{fig:mri_representation_3d} and
\ref{fig:mri_representation_2d}, and $k_z$ is a parameter with $0 \le
k_z/K \le 1/2$.  The eigenvalue $s$ determines the temporal evolution
of the parasitic mode and it must be solved for together with the
amplitudes $\delta{\bb{v}}_0(z)$ and $\delta{\bb{B}}_0(z)$, which are
$2\pi/K$-periodic functions in $z$.  The solutions (\ref{eq:v_sol})
and (\ref{eq:b_sol}) are periodic if $k_z/K$ is a rational number.

The set of six differential Equations~(\ref{eq:euler_pi}) and
(\ref{eq:induction_pi}) can in principle be solved for the secondary
velocity and magnetic fields ($\delta P$ can be eliminated using $\del
\bcdot \delta\bb{v} =0$) by requiring that the ``boundary conditions''
\begin{eqnarray}
\label{eq:bc_v}
\delta \bb{v}(\bb{x} + 2\pi/K \check{\bb{z}},t) &=& \delta \bb{v}(\bb{x},t)
\exp{(2\pi i k_z/K)} \,, \\
\label{eq:bc_b}
\delta \bb{B}(\bb{x} + 2\pi/K \check{\bb{z}},t) &=& \delta \bb{B}(\bb{x},t)
\exp{(2\pi i k_z/K)} \,,
\end{eqnarray}
be satisfied for all $z$.  We follow an alternative procedure that
leads to higher order differential equations for $\delta v_z$ and
$\delta B_z$.  Substituting expressions (\ref{eq:v_sol}) and
(\ref{eq:b_sol}) into Equations~(\ref{eq:euler_pi}) and
(\ref{eq:induction_pi}), using the divergenceless nature of the
perturbed fields, and eliminating the pressure perturbation between
the horizontal and vertical components of Equation (\ref{eq:euler_pi})
we obtain
\begin{eqnarray}
\label{eq:linear_sys_nu_1}
(s + \nu \mathcal{Q})\mathcal{Q}\delta v_z 
- i(\bb{k}_{\rm h}\bcdot\Delta{\bb{v}}) (\mathcal{Q}-K^2) \delta v_z
\nonumber \\
+ i(\bb{k}_{\rm h}\bcdot\Delta{\bb{B}}) (\mathcal{Q}-K^2) \delta B_z
= 0
\,,
\end{eqnarray}
\begin{eqnarray}
\label{eq:linear_sys_eta_1}
(s + \eta \mathcal{Q})\delta B_z 
+ i(\bb{k}_{\rm h}\bcdot\Delta{\bb{B}}) \delta v_z 
- i(\bb{k}_{\rm h}\bcdot\Delta{\bb{v}}) \delta B_z
= 0
\,.\,\,\,
\end{eqnarray}
Here, we have used the explicit form of $\Delta{\bb{v}}$ and
$\Delta{\bb{B}}$ from Equations~(\ref{eq:Delta_V_MRI}) and
(\ref{eq:Delta_B_MRI}), and defined the differential operator
$\mathcal{Q} = k_{\rm h}^2 - \partial_z^2$. These correspond to the
equations of motion for the parasites presented in \citet{PG09}.

Thus, the assumptions stated at the beginning of this Section allows
us to reduce the problem of analyzing the stability of the exact MRI
modes against secondary perturbations in terms of a set of linear,
ordinary differential equations with periodic coefficients.

\subsection{The Eigenvalue Problem and Its Solution}

The set of Equations~(\ref{eq:linear_sys_nu_1}) and
(\ref{eq:linear_sys_eta_1}) can in principle be integrated along the
$z$-coordinate subject to the boundary conditions (\ref{eq:bc_v}) and
(\ref{eq:bc_b}). It is more convenient, however, to work in Fourier
space and transform the differential equations into algebraic
equations, by seeking solutions of the form
\begin{eqnarray}
\label{eq:fourier_v}
\delta v_z(z;k_z) &=& \sum_{n=-\infty}^{\infty} \alpha_n e^{-i(nK+k_z)z}  e^{-i\bb{k}_{\rm h}\bcdot\bb{x}} \,,\\
\label{eq:fourier_b}
\delta B_z(z;k_z) &=& \sum_{n=-\infty}^{\infty}  \beta_n e^{-i(nK+k_z)z} e^{-i\bb{k}_{\rm h}\bcdot\bb{x}} \,.
\end{eqnarray}

When the operator $\mathcal{Q}$ acts on the Fourier Series each
individual terms incurs a factor $Q_n \equiv k_{\rm h}^2 +
(k_z+nK)^2$.  Thus, the differential
Equations~(\ref{eq:linear_sys_nu_1}) and (\ref{eq:linear_sys_eta_1})
lead to recursion relations for the Fourier coefficients
$\{\alpha_n\}$ and $\{\beta_n\}$ of the form
\begin{eqnarray}
  0 &=& 2(s + \nu Q_n) Q_n\alpha_n 
  \nonumber \\
  &-& i \, (\bb{k}_{\rm h} \bcdot \bb{B}_0)  [(Q_{n-1}-K^2) \beta_{n-1}  + (Q_{n+1}-K^2)  \beta_{n+1}] 
  \nonumber \\
  &-&      (\bb{k}_{\rm h} \bcdot \bb{V}_0)  [(Q_{n-1}-K^2)
  \alpha_{n-1} - (Q_{n+1}-K^2) \alpha_{n+1}]  \,, 
  \nonumber \\
  \\
  0 &=& 2(s + \eta Q_n)\beta_n 
  \nonumber \\
  &-& i \, (\bb{k}_{\rm h} \bcdot \bb{B}_0) (\alpha_{n-1}+\alpha_{n+1}) 
  -      (\bb{k}_{\rm h} \bcdot \bb{V}_0) (\beta_{n-1}-\beta_{n+1}) \,.
  \nonumber \\
\end{eqnarray}
It is convenient to use the natural scales provided by the primary MRI
mode and rescale the variables: $Q_n \rightarrow Q_n/K^2$, $k_{\rm h}
\rightarrow k_{\rm h}/K$, $\alpha_n \rightarrow \alpha_n/B_0$,
$\beta_n \rightarrow \beta_n/B_0$. We thus obtain
\begin{eqnarray}
  \frac{s}{KB_0} \alpha_n &=& - \frac{\nu K^2}{KB_0} Q_n \alpha_n  
  \nonumber \\
  &+& i \, \frac{(\bb{k}_{\rm h} \bcdot \check{\bb{B}}_0)}{2Q_n}  [(Q_{n-1}-1) \beta_{n-1}  + (Q_{n+1}-1)  \beta_{n+1}] 
  \nonumber \\
  &+&    \frac{(\bb{k}_{\rm h} \bcdot \check{\bb{V}}_0)}{2Q_n} \frac{V_0}{B_0}  [(Q_{n-1}-1) \alpha_{n-1} - (Q_{n+1}-1) \alpha_{n+1}] \,, 
  \nonumber \\
  \label{eq:eigensystem_v}
\end{eqnarray}
\begin{eqnarray}
 \frac{s}{KB_0} \beta_n = - \frac{\eta K^2}{KB_0} Q_n \beta_n 
+ i \,\frac{(\bb{k}_{\rm h} \bcdot \check{\bb{B}}_0)}{2} (\alpha_{n-1}+\alpha_{n+1})
\nonumber \\
+     \frac{(\bb{k}_{\rm h} \bcdot \check{\bb{V}}_0)}{2}
\frac{V_0}{B_0} (\beta_{n-1}-\beta_{n+1}) \,, \,\,\,
\label{eq:eigensystem_b}
\end{eqnarray}
where the versors $\check{\bb{V}}_0 \equiv \bb{V}_0/V_0$ and
$\check{\bb{B}}_0 \equiv \bb{B}_0/B_0$ provide the direction of the
MRI velocity and magnetic fields. Recall that the ratio of the
velocity to magnetic field amplitudes, $V_0/B_0$, for the primary MRI
mode is not an independent variable; it can be calculated for each set
of values $(\nu, \eta, K)$ \citep{PC08}.  It is convenient to scale
the Fourier coefficients using $B_0$, rather than $V_0$, because the
ratio $V_0/B_0$ is proportional to $\Lambda_\eta$ for $\Lambda_\eta
\ll 1$.

The system of coupled linear Equations~(\ref{eq:eigensystem_v}) and
(\ref{eq:eigensystem_b}), with $n = -\infty, \ldots, \infty$, can be
written in matrix form as $M \bb{q} = s \bb{q}$, where $M$ is a
band-diagonal, complex, non-Hermitian matrix and the components of the
eigenvector $\bb{q}$ are defined according to $q_{n} = \alpha_{n/2}$
for $n=2m$ and $q_{n} = \beta_{n/2+1}$ for $n=2m+1$.  Boundary
conditions such that $(\alpha_n,\beta_n)\rightarrow 0$ must be imposed
on the eigenvectors in order to ensure that the Fourier Series
(\ref{eq:fourier_v}) and (\ref{eq:fourier_b}) converge as
$|n|\rightarrow\infty$.

In practice, we set $\alpha_{\rm N} = \beta_{\rm N} = 0$ for $|N|\ge
N_{\rm max}$ and diagonalize the $2(2N_{\rm max} +1) \times 2(2N_{\rm
  max} +1)$ matrix $M$ for a given set of values $(\nu, \eta, K, B_0,
k_z, \theta, k_{\rm h})$. The algorithm employed to diagonalize the
matrix $M$ is based on the LAPACK routine ZGEEVX, using the option of
applying a balancing transformation to improve the conditioning of the
eigenvalues and eigenvectors. We solve for the finite set of
$2(2N_{\rm max} +1)$ eigenvalues $\{s\}$ and eigenvectors $\{\bb{q}\}$
(if needed), taking increasingly large values of $N_{\rm max}$ until
convergence to the desired accuracy is reached; $N_{\rm max} = 30$
seems to do a very good job across the parameter space explored.

\section{Parasitic Instabilities and Saturation of MRI Modes}
\label{sec:PI_saturation}

It may seem that the large number of independent variables involved
could prevent us from obtaining a global understanding of the
saturation of the MRI across the parameter space. This is not the
case, however. We first present the results of a systematic
calculation of the saturation amplitude of the fastest growing MRI
modes, and their associated stresses, and show that they exhibit
simple behaviors.  In Section~\ref{sec:asymptotia} we provide an
explanation of these results by analyzing the behavior of the primary
MRI modes and the parasitic modes responsible for their saturation in
different regions of parameter space.

\subsection{Parameter Space of Interest}
\label{sec:parameter_space}

Astrophysical disks, MRI laboratory experiments, and
numerical simulations span a wide range of values in the parameter
space defined by $(\Lambda_\nu, \Lambda_\eta)$, or equivalently
$(\Lambda_\nu, {\rm Pm})$:

\emph{Astrophysical Disks}: The temperature and density in an
accretion disk vary by many orders of magnitude across its radius.
Thus, the kinematic viscosity, $\nu$, and magnetic diffusivity,
$\eta$, are expected to span a wide range of values.  In most
astrophysical environments, the kinematic viscosity is usually very
small and thus $\Lambda_\nu \gg 1$. For a protoplanetary disk
$\Lambda_\eta$ is a steep function of radius with $\Lambda_\eta \simeq
5\times 10^{-11} (B/0.1\textrm{G})^2 (R/\textrm{AU})^{37/4}$, where
$R$ is the distance to the central star, see \citet{SM99}.
\citet{BH08} find that the magnetic Prandtl number for accretion disks
around black holes decreases monotonically with increasing radius and
lies in the range ${\rm Pm} \sim 10^{-3}$--$10^3$ for $R/R_{\rm s}
\simeq 3$--$10^{3}$, where $R_s$ is the Schwarzschild radius.

\emph{Laboratory Experiments}: Current Taylor-Couette experiments
using liquid metals to realize the MRI in laboratories are
characterized by physical parameters given by $\nu \sim 3 \times
10^{-3}$ cm$^2$/s, $\eta \sim 3 \times 10^3$cm$^2$/s, $\rho \sim
6$gr/cm$^{-3}$, $\Omega\sim 10$--$40$Hz, and $\bar{B}_z\sim 1$--$4
\times 10^3$G \citep{Nornbergetal10}.  The value of $\Lambda_\nu$ can
be varied by changing the strength of the background magnetic field or
the rotation rate; characteristic values are of the order of
$\Lambda_\nu \sim 10^5$--$10^6$.  The magnetic Prandtl number is
approximately constant with ${\rm Pm} \sim 10^{-6}$.

\emph{Numerical Simulations}: There have been recent numerical
simulations of the non-linear development of the MRI in the
shearing-box approximation exploring the effects of explicit viscosity
and resistivity.  Using our definitions, the range of parameters
explored correspond to $\Lambda_\nu \sim 1$--$100$ and ${\rm Pm} \sim
0.1$--$10$ in the three-dimensional simulations in \citet{LL07} and
$\Lambda_\nu \sim 0.01$--$10$ and ${\rm Pm} \sim 10^{-3}$--$10^4$ in
the axisymmetric simulations in \citet{MS08}.

Motivated by previous studies that suggest that the saturation
amplitude of the MRI is sensitive to the magnetic Prandtl number, see
e.g., \citet{UMR07, URM07} and \citet{LL07}; we consider
$(\Lambda_\nu, {\rm Pm})$ as independent parameters.  In order to
explore in a systematic way the large region of parameter space
relevant to the various environments mentioned above, we consider the
range of parameters defined by
\begin{eqnarray}
\label{eq:parameter_nu}
\Lambda_\nu &=&\{1, 10, \ldots, 10^7\} \\
\label{eq:parameter_Pm}
{\rm Pm} &=& \{10^{-7}, 10^{-6}, \ldots, 10^7\} \,.
\end{eqnarray}
This range of values is large enough to capture all the interesting
behaviors and allows us to derive asymptotic scalings that are
applicable outside this region of parameter space.


\begin{figure}
  \includegraphics[width=\columnwidth,trim=0 0 0 0]{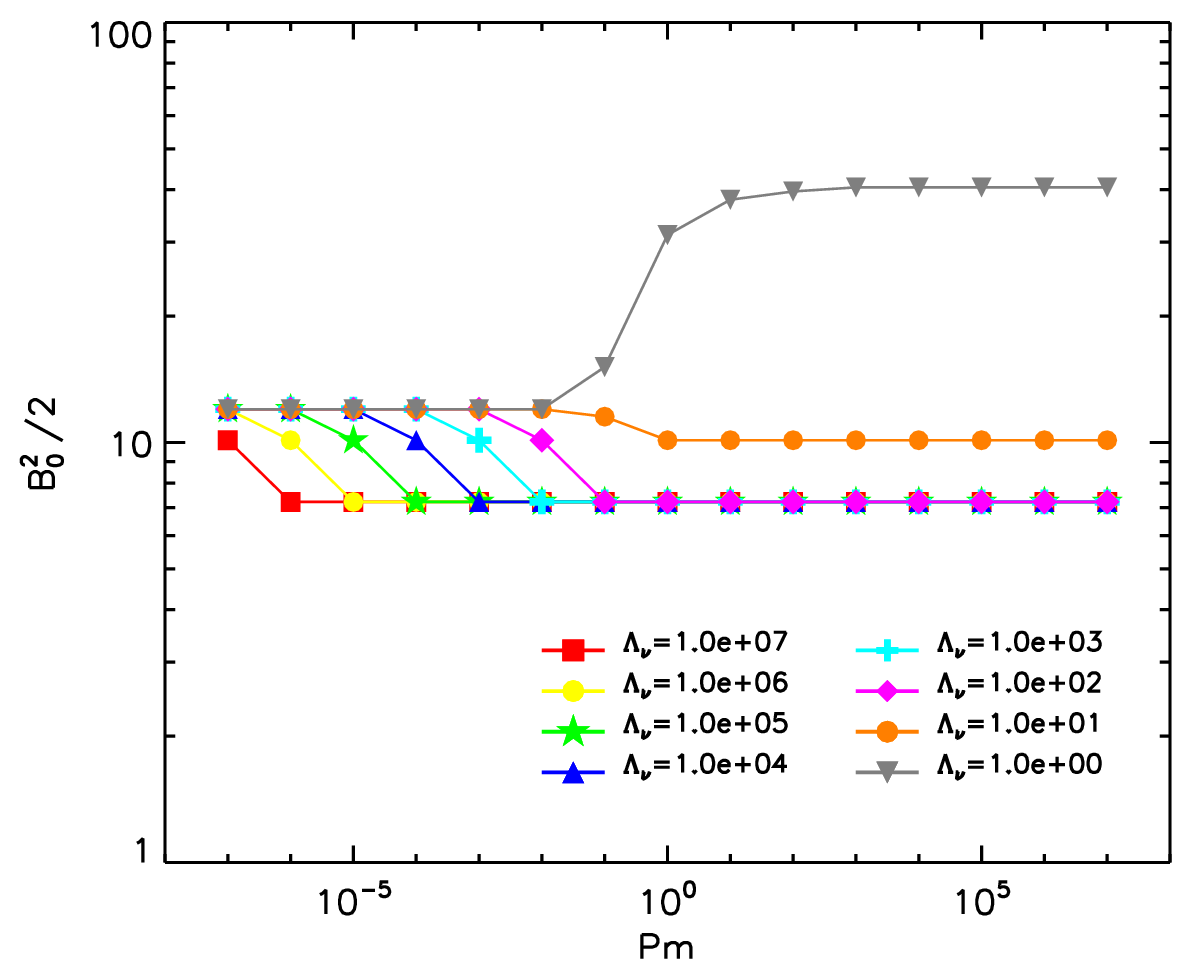}
  \caption{Magnetic energy density corresponding to the fastest
    growing primary MRI mode when the growth rate of the fastest
    parasitic mode matches $\Gamma_{\rm max}(\nu,\eta)$, i.e., $B_0 =
    B_0^{\rm sat}$.  For $\Lambda_\nu \gtrsim 10$, there are two
    clear asymptotic regimes that correspond to ${\rm Pm} \,
    \Lambda_\nu \equiv \Lambda_\eta$ larger or smaller than unity. The
    corresponding modes are associated with Kelvin-Helmholtz and
    tearing modes respectively, see Section~\ref{sec:asymptotia}.}
  \label{fig:B0_sat}
\end{figure}

\subsection{Magnetic Energy Density and MRI-Stresses}

For each pair of values $(\Lambda_\nu, {\rm Pm})$, we analyze the
stability of the fastest growing primary MRI mode $K_{\rm
  max}(\nu,\eta)$ by solving the system of
Equations~(\ref{eq:eigensystem_v}) and (\ref{eq:eigensystem_b}) as a
function of the variables $(B_0, k_z, \theta, k_{\rm h})$.  In order
to explore the parameter space thoroughly, and identify the fastest
parasitic modes, we find the growth rate of the secondary modes for a
grid of values given by $k_z=\{0.0, 0.1, \ldots, 0.5\}$,
$\theta=\{0^{\circ}, 5^{\circ}, \ldots, 180^{\circ}\}$, and $k_{\rm h}
= \{0.0, 0.01, \ldots, 1.0\}$\footnote{The primary MRI modes seem to
  be always stable for $k_{\rm h}>1$.}. We find the fastest growing
parasitic mode for a fixed value of the primary MRI magnetic field
$B_0$ and iterate this procedure increasing $B_0$ until the fastest
secondary instability matches the growth rate of the fastest MRI mode,
i.e., $s_{\rm max}(\nu,\eta,K_{\rm max}) = \Gamma_{\rm
  max}(\nu,\eta)$.  This corresponds to the value $B_0^{\rm
  sat}(\nu,\eta)$ that we denominate the ``saturation'' amplitude of
the MRI magnetic field \citep{PG09}, which is shown in
Figure~\ref{fig:B0_sat}.

For values $\Lambda_\nu \gtrsim 10$, the amplitude $B_0^{\rm sat}$
displays two clear asymptotic behaviors, i.e., $(B_0^{\rm sat})^2/2
\simeq 7$ and $(B_0^{\rm sat})^2/2 \simeq 12$, depending on the
magnetic Prandtl number.  The 'critical' values for this transition,
${\rm Pm}^{\rm c}$, depend on $\Lambda_\nu$ and satisfy
\begin{eqnarray} {\rm Pm}^{\rm c}(\Lambda_\nu) \, \Lambda_\nu \equiv
  \Lambda_\eta^{\rm c} \simeq 1 \quad \textrm{for} \quad \Lambda_\nu \gtrsim
  10 \,.
\end{eqnarray}
Therefore, for $\Lambda_\nu \gtrsim 10$, there is a critical Elsasser
number of order unity that distinguishes two different regimes
characterized by different saturation amplitudes, $B_0^{\rm sat}$.
For $\Lambda_\nu = \{1, 10\}$, the MRI can reach amplitudes $(B_0^{\rm
  sat})^2/2 \simeq 40$ for a wide range of magnetic Prandtl numbers
${\rm Pm} \gtrsim 1$. This behavior is due to the viscous quenching of
Kelvin-Helmholtz modes discussed in \citet{PG09} for ${\rm Pm} \gtrsim
1$.

The saturation amplitude $B_0^{\rm sat}$ varies roughly by only one
order of magnitude across the large parameter space that we consider
and is thus fairly insensitive to the value of the dissipation
coefficients.  In particular, as resistivity increases, i.e.,
$\Lambda_\eta \lesssim 1$, the magnetic field that can be generated by
the MRI reaches an asymptotic, constant value.  The situation is quite
different for the stresses, however.


\begin{figure}
  \includegraphics[width=\columnwidth,trim=0 0 0 0]{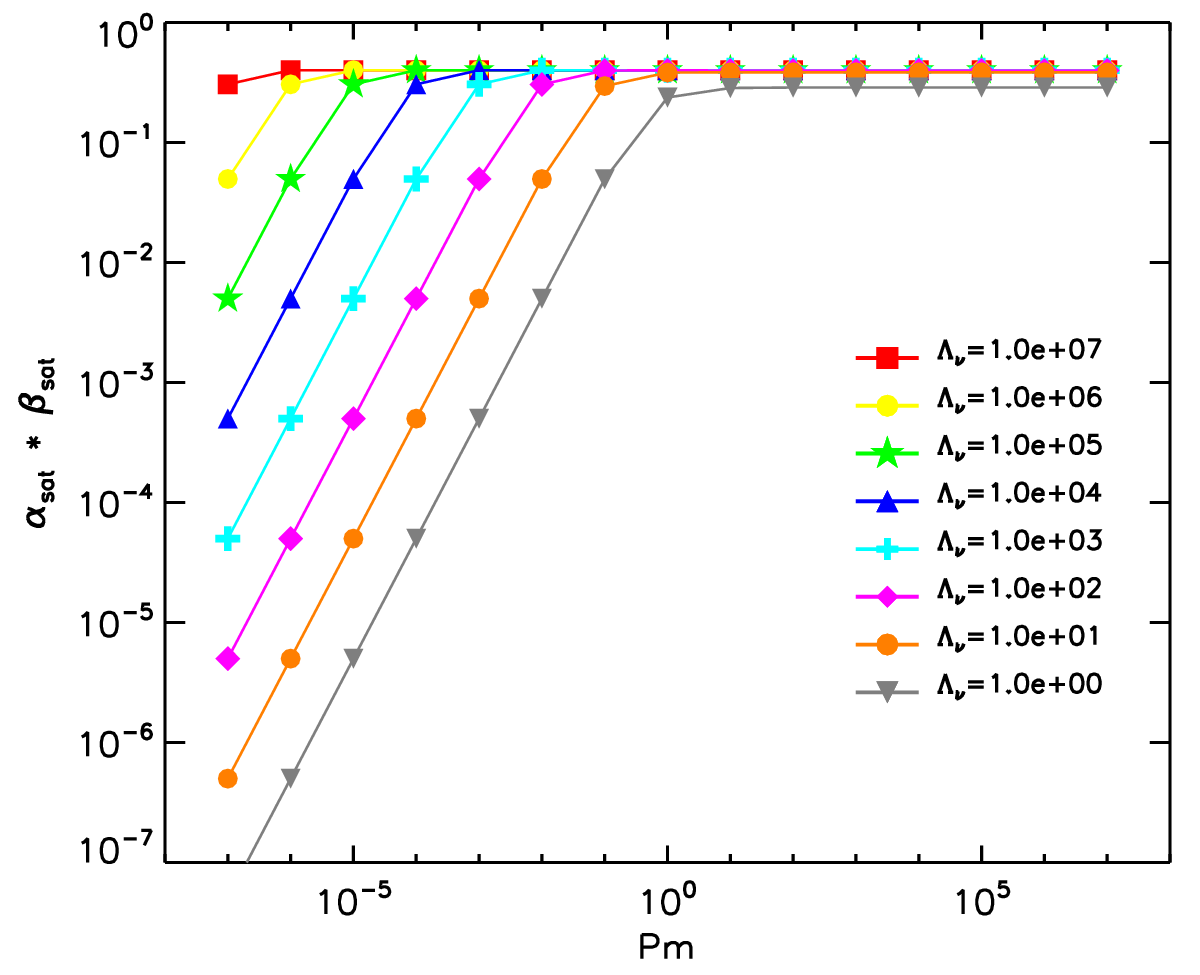}
  \caption{Predicted values for the product $\alpha_{\rm
      sat}\beta_{\rm sat}$ if saturation occurs when the fastest
    parasitic and primary MRI growth rates match.  In the limit,
    $\Lambda_\nu, \, {\rm Pm} \gg 1$, $\alpha_{\rm sat}\beta_{\rm
      sat}\rightarrow 0.4$. In the inviscid, resistive limit, i.e.,
    $\Lambda_\nu \gg 1$ and ${\rm Pm} \ll 1$, $\alpha_{\rm
      sat}\beta_{\rm sat} \rightarrow 0.5 \, {\rm Pm} \, \Lambda_\nu$.
    Despite the fact that the magnetic field at saturation asymptotes
    to a constant value, see Fig.~\ref{fig:B0_sat}, the dimensionless
    stress decreases linearly with ${\rm Pm} \, \Lambda_\nu \equiv
    \Lambda_\eta$ for ${\rm Pm} \,\Lambda_\nu \equiv \Lambda_\eta
    \lesssim 1$.}
  \label{fig:alpha_beta_sat}
\vspace{2.5mm}
\end{figure}


\begin{figure*}
  \includegraphics[width=2\columnwidth,trim=0 0 0 0]{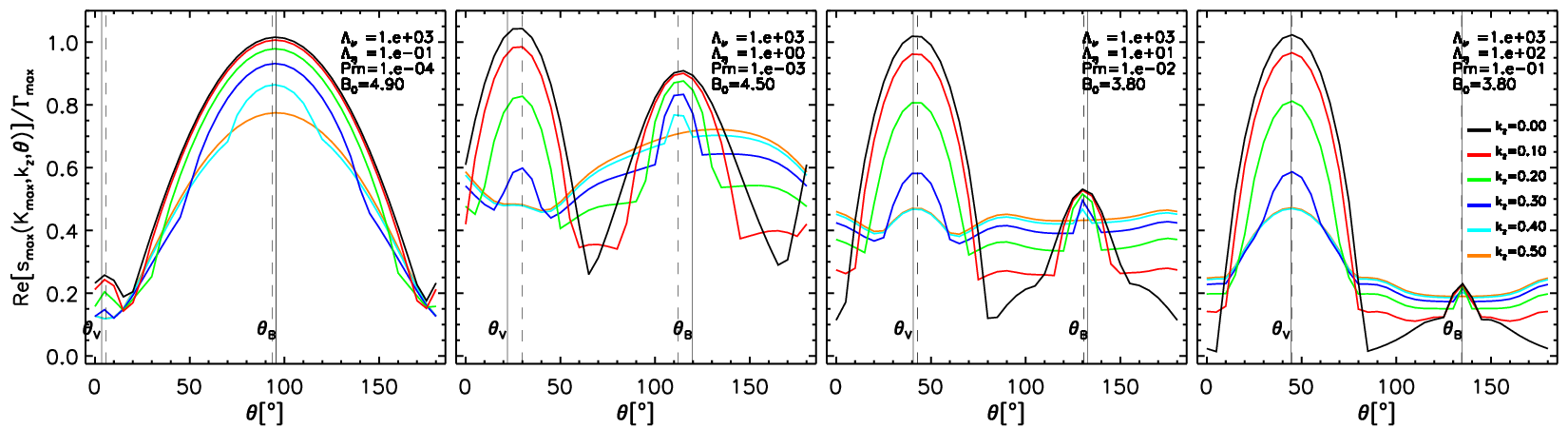}
  \includegraphics[width=2\columnwidth,trim=0 0 0 0]{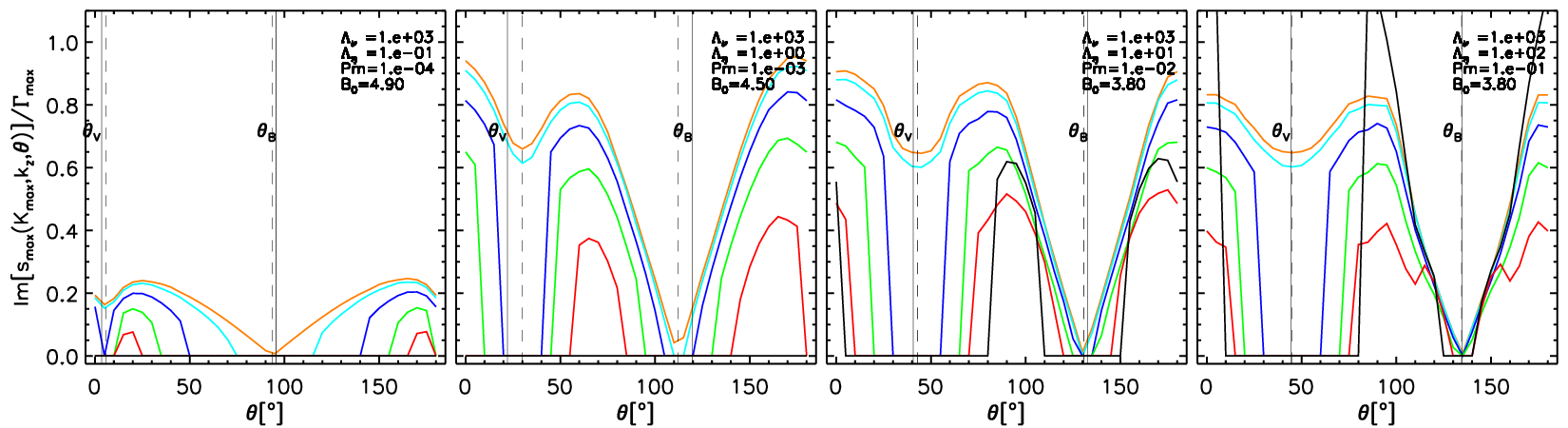}
  \caption{Real and imaginary parts for the fastest parasitic
    instabilities normalized by the growth rate of the fastest MRI
    mode as a function of the orientation of the horizontal wavevector
    $\bb{k}_{\rm h}$ with respect to the radial ($\theta=0$)
    direction. The panels correspond to $\Lambda_\eta = \{0.1, 1, 10,
    10^2\}$, from left to right, with $\Lambda_\nu \gg1$.  The MRI
    magnetic field $B_0=B_0^{\rm sat}(\nu,\eta)$ is such that the
    fastest parasitic growth rate, maximized over $k_{\rm h}$,
    $\theta$, and $k_z$, matches the growth rate of the fastest
    primary MRI mode, $\Gamma_{\rm max}(\nu,\eta)$.  The parasites
    present fastest growth along the directions $\theta_{\rm
      max}\approx\theta_{\rm V}$ and $\theta_{\rm
      max}\approx\theta_{\rm B}$, shown as solid lines, and they are
    associated with Kelvin-Helmholtz and tearing mode instabilities,
    respectively.}
  \label{fig:s_max_pi}
\end{figure*}

The dimensionless stress at saturation is given by $\alpha_{\rm sat}
\equiv \bar{T}^{\rm sat}_{r\phi}/(L_z\Omega_0)^2$, where $\bar{T}^{\rm
  sat}_{r\phi}\equiv \bar{R}^{\rm sat}_{r\phi}-\bar{M}^{\rm
  sat}_{r\phi}$, is the sum of the Reynolds and Maxwell stresses
\begin{eqnarray}
\bar{R}^{\rm sat}_{r\phi} &\equiv&\frac{1}{L_z}\int_{-L_z\!/2}^{L_z\!/2} 
V^{\rm sat}_{0,r}(z)V^{\rm sat}_{0,\phi}(z) dz \,, \\
\bar{M}^{\rm sat}_{r\phi} &\equiv&\frac{1}{L_z}\int_{-L_z\!/2}^{L_z\!/2} 
B^{\rm sat}_{0,r}(z)B^{\rm sat}_{0,\phi}(z) dz \,.
\end{eqnarray}
Integrating these expressions we obtain the dimensionless stress
$\alpha_{\rm sat}$ in terms of the parameter\footnote{\citet{PG09}
  defined $\alpha_{\rm sat} \equiv \bar{T}^{\rm sat}_{r\phi}/(SL_z)^2$
  and $\beta \equiv S^2 L_z^2/\bar{v}_{{\rm A}z}^2$, with
  $S\equiv3\Omega_0/2$, in order to compare results with \citet{LL07}.
  The definitions adopted here do not affect the expression or the
  numerical value for the product $\alpha\beta$.} $\beta \equiv
\Omega_0^2 L_z^2/\bar{v}_{{\rm A}z}^2$,
\begin{eqnarray}
\alpha_{\rm sat} &=& \frac{1}{2\beta} 
[(V_0^{\rm sat})^2 \sin\theta_{\rm V} \cos\theta_{\rm V} -
( B_0^{\rm sat})^2 \sin\theta_{\rm B} \cos\theta_{\rm B}]
\,. \nonumber \\ 
\end{eqnarray}

We can obtain an expression for the ratio between the stress and the
magnetic energy density associated with the primary MRI mode by relating
the initial value of the $\beta$-parameter, $\beta \propto
1/\bar{B}_z^2$, with an estimate for its corresponding value at
saturation, i.e., $\beta_{\rm sat} = \beta/B_0^2$. This leads to
\begin{eqnarray}
\alpha_{\rm sat} \beta_{\rm sat} &=& \frac{1}{2} 
\left[\left(\frac{V_0^{\rm sat}}{B_0^{\rm sat}}\right)^2 \sin\theta_{\rm V} \cos\theta_{\rm V} -
  \sin\theta_{\rm B} \cos\theta_{\rm B}\right] \,. \nonumber \\ 
\label{eq:alpha_beta}
\end{eqnarray}
The dependence of the product $\alpha_{\rm sat} \beta_{\rm sat}$ on
${\rm Pm}$ for different values of $\Lambda_\nu$ is shown in
Figure~\ref{fig:alpha_beta_sat}. The existence of two different
asymptotic behaviors for $\Lambda_\nu \gtrsim 10$ is evident:
\begin{eqnarray}
\label{eq:alpha_beta_sat_ideal}
  \alpha_{\rm sat}\beta_{\rm sat} &\simeq& 0.4 \quad \quad \quad
  \quad \textrm{for} \quad {\rm Pm} \, \Lambda_\nu > 1 \,, \\
\label{eq:alpha_beta_sat_resis}
  \alpha_{\rm sat}\beta_{\rm sat} &\simeq& 0.5 \, {\rm Pm} \, \Lambda_\nu  \quad \textrm{for} \quad {\rm Pm} \, \Lambda_\nu < 1 \,.
\end{eqnarray}

The asymptotic behavior for $\Lambda_\nu \gg 1$ and ${\rm Pm \gg 1}$
corresponds to the ideal MHD limit with $\Lambda_\eta \gg \Lambda_\nu
\gg 1$.  It is worth mentioning that numerical simulations of MRI
driven turbulence (with no explicit dissipation), carried out over a
wide range of physical conditions, lead to saturation values for the
parameters $\alpha_{\rm sat}$ and $\beta_{\rm sat}$ that vary by
several orders of magnitude \citep[see, e.g.][and references
therein]{PCP06a}. However, their product remains roughly constant with
$\alpha_{\rm sat}\,\beta_{\rm sat} \simeq 0.5$
\citep{HGB95,Sanoetal04}. Despite the approximations that we adopted
for calculating this quantity, the value in
Equation~(\ref{eq:alpha_beta_sat_ideal}) is remarkably close to the
results obtained in numerical simulations. We note that \citet{BPV08}
provide independent arguments, involving turbulent eddies, to support
the idea that $\alpha_{\rm sat}$ and $\beta_{\rm sat}$ should be
anti-correlated. They estimate that $\alpha_{\rm sat}\,\beta_{\rm sat}
\simeq 0.7$, for an isothermal equation of state.

The product $\alpha_{\rm sat}\beta_{\rm sat}$ presents a very
different behavior below a 'critical' value, ${\rm Pm}^{\rm
  c}(\Lambda_\nu) \, \Lambda_\nu \equiv \Lambda_\eta^{\rm c} \simeq
1$.  In this regime, the stress decreases linearly with both ${\rm
  Pm}$ and $\Lambda_\nu$ for ${\rm Pm} \, \Lambda_\nu \lesssim 1$.
The product ${\rm Pm} \, \Lambda_\nu$ corresponds of course to the
Elsasser number, $\Lambda_\eta$.  We write the expression for ${\rm
  Pm} \, \Lambda_\nu$ in Equation~(\ref{eq:alpha_beta_sat_resis})
explicitly because it is not a priori evident that the quantity
$\alpha_{\rm sat} \beta_{\rm sat}$ should scale linearly with both
${\rm Pm}$ and $\Lambda_\nu$. For instance, it could have depended on
some power of the magnetic Prandtl number, e.g., ${\rm Pm}^{1/2}$ with
a weaker dependence on $\Lambda_\nu$. This is indeed the type of
dependence observed in the numerical simulations in \citet{LL07},
which address the regime $\Lambda_\nu \sim 1$--$100$ and ${\rm Pm}
\sim 0.1$--$10$.  However, in the regime $\Lambda_\nu \gtrsim 10$, our
calculations lead to MRI stresses that depend only on the Elsasser
number $\Lambda_\eta$ with $\bar{T}_{r\phi} \propto \Lambda_\eta$ for
$\Lambda_\eta \lesssim 1$.

The existence of a critical Elsasser number of order unity is related
to the fact that the fastest growing secondary modes are associated
with Kelvin-Helmholtz instabilities for $\Lambda_\eta > 1$ and tearing
instabilities for $\Lambda_\eta < 1$, provided that $\Lambda_\nu
\gtrsim 10$.  In order to provide support to these assertions, it is
useful to examine the behavior of the growth rates of the parasitic
modes as a function of the direction of the horizontal wavevector
$\bb{k}_{\rm h}$, i.e., the angle $\theta$, see
Figures~\ref{fig:mri_representation_3d} and
\ref{fig:mri_representation_2d}.


\begin{deluxetable*}{cccrrccccc}[t]
  \tablewidth{0pc} 
\tablecaption{Characterization of Fastest MRI and Parasitic Modes
\label{table:mri_pi_mode_data}}
\tablehead{
\colhead{$\Lambda_\eta$}&
\colhead{$\Gamma_{\rm max}/\Omega_0, s_{\rm max}/\Omega_0$}&
\colhead{$K_{\rm max}$}&
\colhead{$\theta_{\rm V}[^\circ]$}&
\colhead{$\theta_{\rm B}[^\circ]$}&
\colhead{$\theta_{\rm max}[^\circ]$}&
\colhead{$k_{\rm h, max}/K_{\rm max}$}&
\colhead{$B_0^{\rm sat}/\bar{B}_z$}&
\colhead{$V_0/B_0$} &
\colhead{Type}
}
\startdata
1.E$-$03 & 7.50E$-$04 & 8.66E$-$04 &  0.0 &  90.1 & 90.0 & 0.48 & 4.9 & 0.0017 & TM \\
1.E$-$02 & 7.50E$-$03 & 8.66E$-$03 &  0.0 &  90.6 & 90.0 & 0.48 & 4.9 & 0.0173 & TM \\
1.E$-$01 & 7.39E$-$02 & 8.55E$-$02 &  3.5 &  95.6 & 95.0 & 0.48 & 4.9 & 0.1689 & TM \\
1.E$+$00 & 4.28E$-$01 & 5.16E$-$01 & 22.1 & 119.7 & 30.0 & 0.59 & 4.5 & 0.7187 & KH \\
1.E$+$01 & 6.96E$-$01 & 8.88E$-$01 & 40.6 & 132.9 & 40.0 & 0.59 & 3.8 & 0.7816 & KH \\
1.E$+$02 & 7.44E$-$01 & 9.59E$-$01 & 44.5 & 134.7 & 45.0 & 0.59 & 3.8 & 0.7751 & KH \\
1.E$+$03 & 7.49E$-$01 & 9.67E$-$01 & 45.0 & 135.0 & 45.0 & 0.59 & 3.8 & 0.7743 & KH \\
1.E$+$04 & 7.50E$-$01 & 9.67E$-$01 & 45.0 & 135.0 & 45.0 & 0.59 & 3.8 & 0.7742 & KH
\enddata
\tablecomments{Data corresponding to the fastest growing MRI modes and
  their associated fastest parasitic modes for $\Lambda_\nu \gg 1$.
  Note the asymptotic behaviors at low and high Elsasser numbers and
  the sharp transition around $\Lambda_\eta \simeq 1$.  The resistive
  ($\Lambda_\eta \ll 1$) and ideal ($\Lambda_\eta \gg 1$) MHD regime
  are dominated by tearing modes (TM) and Kelvin-Helmholtz (KH) modes,
  respectively.}
\end{deluxetable*}

\subsection{Kelvin-Helmholtz vs. Tearing Modes}

It is of particular interest to understand which type of secondary
modes are the first to reach growth rates that are comparable to the
growth rate of the MRI in different regions of the parameter spaced
spanned by the dissipation coefficients.

The upper and lower sets of panels in Figure~\ref{fig:s_max_pi} show
the real and imaginary parts of the eigenvalues, $s_{\rm
  max}(\nu,\eta, K_{\rm max}, k_z, \theta)$, corresponding to the
fastest growing parasitic modes that feed off the fastest primary MRI
mode for $B_0 = B_0^{\rm sat}(\nu,\eta)$.  The different panels in
this figure correspond to $\Lambda_\eta=\{0.1, 1, 10^{1},
10^{2}\}$, from left to right, with $\Lambda_\nu=10^3$.  The various
curves in each panel correspond to different values of the parameter
$k_z$. The vertical solid lines show the angles $\theta_{\rm
  V}(\nu,\eta)$ and $\theta_{\rm B}(\nu,\eta)$ associated with the
velocity and magnetic fields of the fastest MRI mode, see
Figures~\ref{fig:mri_representation_3d} and
~\ref{fig:mri_representation_2d}.  The vertical dashed lines in each
panel show the directions perpendicular to $\theta_{\rm B}$ (left) and
$\theta_{\rm V}$ (right).  The MRI velocity and magnetic fields are
close to orthogonal, except for $\Lambda_\eta \simeq 1$, where the
directions parallel to $\theta_{\rm V}$ and perpendicular to
$\theta_{\rm B}$ differ by a few degrees.

The fastest parasitic modes that determine the values of $B_0^{\rm
  sat}$ shown in Figure~\ref{fig:B0_sat} are non-axisymmetric
($\theta_{\rm max} \ne 0$); their horizontal wavevectors, $\bb{k}_{\rm
  h, max}$, are nearly aligned with either the velocity or the
magnetic field of the primary mode.  The first type reach their
maximum growth rates for $\theta_{\rm max} \simeq \theta_{\rm V}$,
i.e., the direction of the MRI velocity field, $\Delta \bb{V}$. These
correspond to Kelvin-Helmholtz modes and dominate for $\Lambda_\eta
\ge 1$.  The second type of modes grow the fastest along the direction
$\theta_{\rm max} \simeq \theta_{\rm B}$, i.e., the direction of the
MRI magnetic field, $\Delta \bb{B}$. These are tearing modes enabled
by resistive reconnection of the MRI magnetic field. They dominate for
$\Lambda_\eta < 1$ and become increasingly relevant as
$\Lambda_\eta$ decreases.

Independently of the value of the dissipation coefficients, the
fastest growing Kelvin-Helmholtz and tearing modes have the same
vertical periodicity as the primary mode, i.e., $k_z = 0$.  As
illustrated in the lower panels of Figure~\ref{fig:s_max_pi}, the
fastest growing Kelvin-Helmholtz modes have purely real growth rates
for $k_z \lesssim 0.25$; while they have imaginary parts that are
similar to their real parts for $k_z \gtrsim 0.25$. The latter
correspond to the ``Type-II'' modes discussed in \citet{GX94}, see
also \citet{LLB09}.  These are clearly Kelvin-Helmholtz modes since
they present their maximum growth along the direction of the MRI
velocity field.  These modes might be relevant in axisymmetric
simulations of viscous, resistive MRI because they have growth rates
that are comparable, or even larger, than the axisymmetric modes with
$k_z=0$. The fastest growing tearing modes have zero, or very small,
imaginary parts for every value of $k_z$.

It is clear from Figures~\ref{fig:B0_sat}, \ref{fig:alpha_beta_sat},
and \ref{fig:s_max_pi} that there is a sharp transition in behavior
around a critical Elsasser number of order unity.
Table~\ref{table:mri_pi_mode_data} provides information concerning the
growth rates and geometric structure of both the fastest growing
primary MRI mode and the associated fastest parasitic mode for a range
of Elsasser numbers $\Lambda_\eta=\{10^{-3}, \ldots, 10^{4}\}$, with
$\Lambda_\nu=10^{3}$.  Note that the horizontal wavelength of the
fastest growing modes, for both tearing and Kelvin-Helmholtz modes, is
roughly a factor of two larger than the vertical wavelength of the
primary mode upon which they feed.  Tearing modes and Kelvin-Helmholtz
modes present the fastest growth at low and high Elsasser numbers,
respectively.  These two regimes are characterized by asymptotic
behaviors that are already evident for $\Lambda_\eta \simeq 10^{\pm
  2}$.

The existence of these scalings simplifies the characterization of the
physical nature of the parasitic modes in different regions of
parameter space. Thus, we first analyze the asymptotic regimes al
large and low Elsasser number and postpone the explicit analysis of
the physical structure of the eigenmodes until
Section~\ref{sec:eigenmodes}. It will then be evident that it is only
necessary to understand the nature of the fastest modes with moderate
Elsasser number $\Lambda_{\eta} \simeq 1$ and that the characteristics
of the modes for $\Lambda_\eta \ll 1$ and $\Lambda_\eta \gg 1$ can be
obtained using the scaling relations derived next.

\section{Asymptotic Behaviors and Scaling Laws}
\label{sec:asymptotia}

The regime where viscosity plays an important role, i.e., $\Lambda_\nu
\lesssim 10$, was explored in detail in \cite{PG09}.  Hereafter, we
focus our attention on the regime where $\Lambda_\nu \ge 10^{2}$,
i.e., $\Lambda_\nu \gg 1$, which is relevant to astrophysical and
experimental environments.  Here, we analyze the asymptotic behavior
of both primary MRI and secondary modes in this regime and provide
explanations for the behaviors observed in Figures~\ref{fig:B0_sat}
and \ref{fig:alpha_beta_sat}.

\subsection{Asymptotic Behavior of Primary MRI Modes}

In order to better appreciate the asymptotic behavior of the parasitic
instabilities, we need to consider the asymptotic behavior of the
primary MRI modes.  The quantities that define the structure of these
modes, as well as some relationships between various timescales and
amplitudes, are of particular importance and we briefly summarize them
here. We focus our attention in the case of a Keplerian shear profile,
i.e., $q=3/2$.

In ideal MHD, i.e., $\Lambda_\nu, \Lambda_\eta \gg 1$, the maximum
growth achieved by the MRI is $\Gamma_{\rm max} = (3/4) \, \Omega_0$,
and corresponds to the mode with wavenumber $K_{\rm max} =
\sqrt{15/16} \, \Omega_0/\bar{v}_{{\rm A}z}$.  The fastest growth rate
is related to the Alfv\'en frequency, $\omega_{{\rm A}z} = K_{\rm max}
\, \bar{v}_{{\rm A}z}$, associated with the background magnetic field
$\bar{B}_z$, via $\Gamma_{\rm max} = \sqrt{3/5} \, \omega_{{\rm A}z}$
with $\omega_{{\rm A}z} = \sqrt{15/16} \, \Omega_0$.  The magnetic
field and velocity field amplitudes for the fastest primary MRI mode
are related via $V_0 = \sqrt{3/5} \, B_0/\sqrt{4\pi\rho}$
\citep[][]{PCP06a}.  The angles characterizing the planes which
contain the velocity and magnetic fields of the fastest growing MRI
mode are given by $\sin \theta_{\rm V} = - \cos \theta_{\rm B} =
1/\sqrt{2}$.

In the inviscid, resistive MHD limit, i.e., $\Lambda_\nu \gg 1$ and
$\Lambda_\eta \ll 1$, the fastest MRI growth rate $\Gamma_{\rm max} =
(3/4) \, \Lambda_\eta \, \Omega_0$, corresponds to the mode with
wavenumber $K_{\rm max} = \sqrt{3/4} \, \Lambda_\eta \,
(\Omega_0/\bar{v}_{{\rm A}z})$.  In this case, the various relevant
inverse timescales are related via $\Gamma_{\rm max} = \sqrt{3/4} \,
\omega_{{\rm A}z} = \omega_\eta$, where $\omega_{{\rm A}z} = K_{\rm
  max} \, \bar{v}_{{\rm A}z} = \sqrt{3/4} \, \Lambda_\eta \, \Omega_0$
is the Alfv\'en frequency and $\omega_\eta = \eta K_{\rm max}^2 =
(3/4) \, \Lambda_\eta \, \Omega_0$ is the inverse of the resistive
timescale across a lengthscale of the order of the MRI wavelength,
i.e., $K_{\rm max}^{-1}$.  The magnetic field and velocity field
amplitudes for the fastest primary MRI mode are related via $V_0 =
\sqrt{3} \, \Lambda_\eta \, B_0/\sqrt{4\pi\rho}$ \citep{PC08}, and
thus $V_0/B_0 \rightarrow 0$ in the limit $\Lambda_\eta \ll 1$.  The
planes containing the fastest growing MRI velocity and magnetic fields
are characterized by the angles $\sin \theta_{\rm V} = (5/8)
\Lambda_\eta$ and $\cos \theta_{\rm B} = - \Lambda_\eta$.

\subsection{Asymptotic Behavior of Parasitic Growth Rates}

\begin{figure}
  \includegraphics[width=\columnwidth,trim=0 0 0 0]{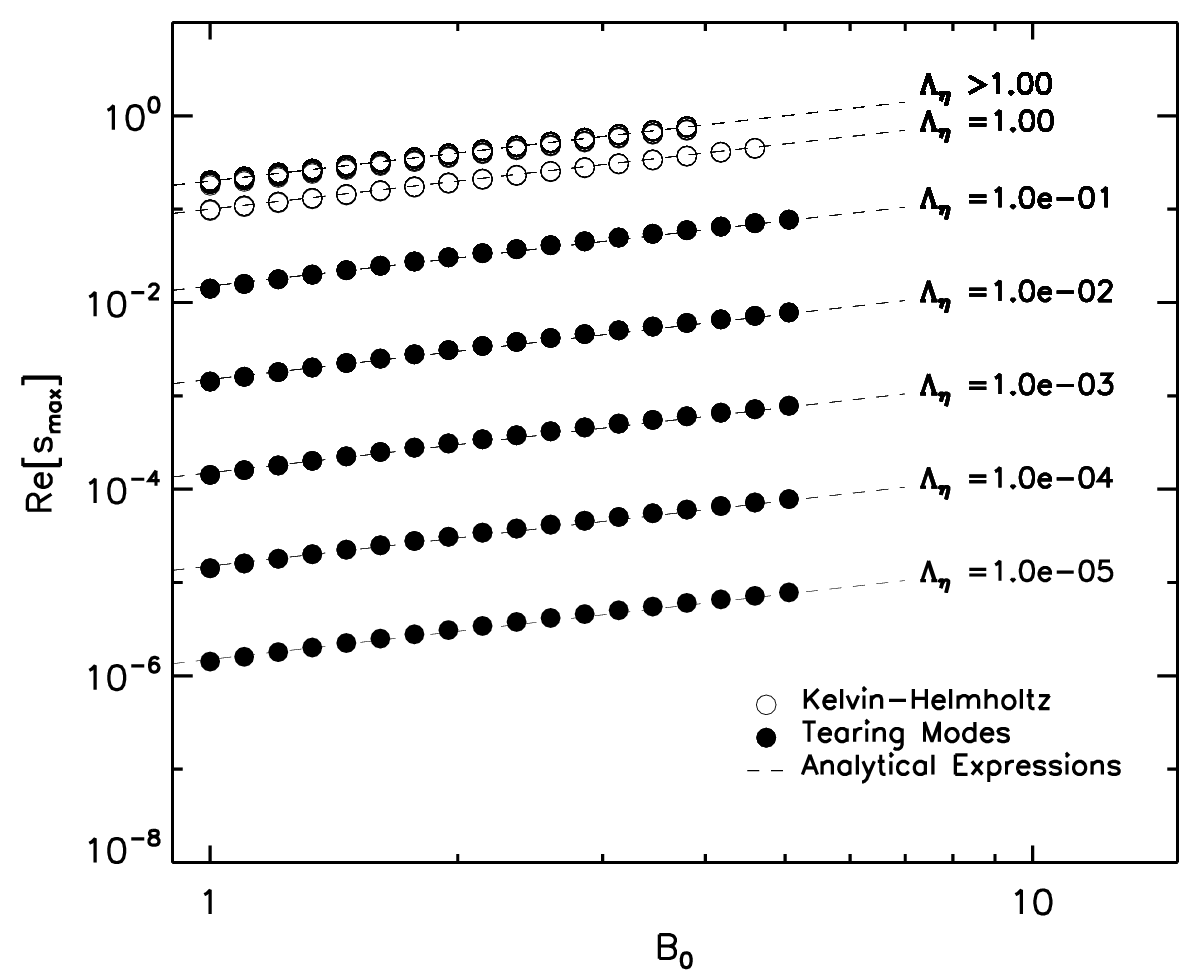}
  \caption{Growth rates for the fastest growing parasitic
    instabilities as a function of the amplitude of the MRI magnetic
    field $B_0$.  The values of $\Lambda_\eta$ associated with each
    set of data points is shown on the right.  Open circles represent
    Kelvin-Helmholtz modes, which exhibit the fastest growing rates
    for $\Lambda_\eta \ge 1$.  Filled circles represent tearing modes,
    which exhibit the fastest growing rates for $\Lambda_\eta < 1$.
    In all cases, the data points are drawn for $B_0 \le B_0^{\rm
      sat}(\nu,\eta)$. The dashed lines, proportional to $B_0$,
    correspond to the analytical expressions in
    Equations~(\ref{eq:s_max_KH_ideal})--(\ref{eq:s_max_TM}).}
  \label{fig:s_max_pi_scaling}
\end{figure}

The dependence of the growth rate of the fastest parasitic modes on
the amplitude of the primary MRI magnetic field, $B_0$, is shown in
Figure \ref{fig:s_max_pi_scaling}.  The open circles identify
Kelvin-Helmholtz modes that feed off the MRI velocity field at $\theta
\simeq \theta_{\rm V}$, which exhibit the fastest growing rates for
$\Lambda_\eta \ge 1$. The filled circles correspond to tearing modes,
that feed off the MRI currents at $\theta \simeq \theta_{\rm B}$.
These are the fastest growing secondary instabilities for
$\Lambda_\eta < 1$.  The following equations\footnote{Although not
  shown in Fig.~\ref{fig:s_max_pi_scaling}, the growth rates of the
  (sub-dominant) Kelvin-Helmholtz modes are given by $s_{\rm max} =
  0.37 B_0 \Lambda_\eta^2$ for $\theta \simeq \theta_{\rm V}, \
  \Lambda_\eta \le 1$.  There does not seem to be simple scaling
  relations for the (sub-dominant) tearing modes in the regime
  $\Lambda_\nu \gg 1$ and $\Lambda_\eta > 1$ for $\theta \simeq
  \theta_{\rm B}$.}, shown as dashed lines in
Figure~\ref{fig:s_max_pi_scaling}, provide an excellent description of
the fastest growth rates associated with Kelvin-Helmholtz modes
\begin{eqnarray}
\label{eq:s_max_KH_ideal}
s_{\rm max} &=& 0.20 B_0  
\quad \quad \textrm{for} \quad 
\theta \simeq \theta_{\rm V}, \, \Lambda_\eta > 1 \,,\\
\label{eq:s_max_KH_non_ideal}
s_{\rm max} &=& 0.10 B_0
\quad \quad \textrm{for} \quad 
\theta \simeq \theta_{\rm V}, \, \Lambda_\eta = 1 \,,
\end{eqnarray}
while the growth rates of the tearing-modes is described by
\begin{eqnarray}
s_{\rm max} &=& 0.15 B_0 \Lambda_\eta 
\quad \textrm{for} \quad 
\theta \simeq \theta_{\rm B}, \, \Lambda_\eta \le 1 \,.
\label{eq:s_max_TM}
\end{eqnarray}

The critical value of the Elsasser number at which the fastest growing
tearing modes (along the direction $\theta \simeq \theta_{\rm B}$)
grow faster than the fastest growing Kelvin-Helmholtz modes (along the
direction $\theta \simeq \theta_{\rm V}$) is very close to (but
slightly less than) unity, i.e., $\Lambda_\eta^{\rm c} \lesssim 1$,
see Figure~\ref{fig:s_max_pi}.  The existence of a an order unity
critical Elsasser number that dictates the linear evolution of the MRI
has already been appreciated on both numerical \citep{SIM98, FSH00}
and analytical grounds \citep[see, e.g.,][]{SM99, PC08}.  Here we
posit that this critical Elsasser number \emph{also} distinguishes
which type of parasitic instability dominates the subsequent evolution
of the MRI.  In the remainder of this Section we argue, and
demonstrate in Section~\ref{sec:eigenmodes}, that there are two
regimes such that the fastest growing parasitic modes correspond to
\begin{eqnarray}
\Lambda_\eta > \Lambda_\eta^{\rm c} \simeq 1 \quad &\rightarrow& \quad \textrm{Kelvin-Helmholtz} \,, \\
\Lambda_\eta < \Lambda_\eta^{\rm c}  \simeq 1 \quad &\rightarrow& \quad \textrm{Tearing Modes} \,.
\end{eqnarray}

The linear dependence of the fastest parasitic growth rates on the
amplitude $B_0$, or $V_0$, when both viscous and resistive effects are
unimportant is to be expected, see Equations~(\ref{eq:eigensystem_v})
and (\ref{eq:eigensystem_b}). However, the linear dependence of the
fastest growth rates on $B_0$ and $\Lambda_\eta$ for $\Lambda_\eta \le
1$ is not a trivial result. These dependencies are responsible for the
asymptotic behaviors described in Section~\ref{sec:PI_saturation}.

\subsubsection{Kelvin-Helmholtz Modes}

For the fastest growing Kelvin-Helmholtz instabilities with
$\Lambda_\nu \gg 1$ and $\Lambda_\eta > 1$, the ratio of the
horizontal parasitic wavenumber to the wavenumber of the fastest
growing primary MRI mode depends very weakly on either $\Lambda_\nu$
or $\Lambda_\eta$, see Table~\ref{table:mri_pi_mode_data}, with
\begin{eqnarray}
\frac{k_{\rm h, max}}{K_{\rm max}} \simeq 0.59
\quad \textrm{for} \quad \Lambda_\nu \gg 1, \Lambda_\eta > 1 \,.
\label{eq:k_to_K_KH_ideal}
\end{eqnarray}
Therefore, restoring the physical dimensions into
Equation~(\ref{eq:s_max_KH_ideal}) we obtain
\begin{eqnarray} 
  s_{\rm max} 
  &\simeq& 0.44 \, \frac{\omega_{{\rm KH},0} \, \Omega_0} {\omega_{{\rm A}z}} 
   \simeq  0.45 \, \omega_{{\rm KH},0} \,,
\label{eq:s_max_KH_ideal_phys}
\end{eqnarray}
where $\omega_{{\rm A}z}$ is the ideal Alfv\'en frequency and
$\omega_{{\rm KH},0} \equiv k_{\rm h} \, V_0$, which corresponds to
the growth rate associated with a Kelvin-Helmholtz instability feeding
off a velocity field discontinuity of amplitude $V_0$.  Therefore, the
amplitude of the MRI magnetic field at which the growth rate of the
parasites matches the growth of the primary mode can be obtained using
(\ref{eq:s_max_KH_ideal}), which leads, in agreement with
Figure~\ref{fig:B0_sat}, to
\begin{eqnarray}
B_0^{\rm sat} \simeq 3.8 \quad \textrm{for} \quad \Lambda_\nu \gg 1,
\Lambda_\eta > 1 \,.
\label{eq:B0_sat_ideal}
\end{eqnarray}


\subsubsection{Tearing Modes}

For the fastest growing tearing modes with $\Lambda_\nu \gg 1$ and
$\Lambda_\eta < 1$, the ratio of the horizontal parasitic wavenumber
to the wavenumber of the fastest growing primary MRI mode depends very
weakly on either $\Lambda_\nu$ or $\Lambda_\eta$, see
Table~\ref{table:mri_pi_mode_data},
\begin{eqnarray}
\frac{k_{\rm h, max}}{K_{\rm max}} \simeq 0.48 
\quad \textrm{for} \quad \Lambda_\nu \gg 1, \Lambda_\eta < 1 \,.
\label{eq:k_to_K_TM}
\end{eqnarray}
Therefore, Equation~(\ref{eq:s_max_TM}) leads to
\begin{eqnarray}
  s_{\rm max} &\simeq& 0.31 \,
  \frac{\omega_{{\rm A}0} \, \omega_{{\rm A}z}} {\omega_\eta} \simeq
  0.36 \, \omega_{{\rm A}0} \,,
\label{eq:s_max_TM_phys}
\end{eqnarray}
where we have defined $\omega_{{\rm A}0} \equiv k_{\rm h}
B_0/\sqrt{4\pi\rho}$, as the Alfv\'en frequency associated with the
horizontal MRI magnetic field $B_0$ and used that $\omega_\eta =
\sqrt{3/4} \, \omega_{{\rm A}z}$. The amplitude of the magnetic field
that the MRI needs to grow to in order for the growth rate of the
parasites to be as large as that corresponding to the primary mode is
obtained using Equation~(\ref{eq:s_max_TM}). This leads, in agreement
with Figure~\ref{fig:B0_sat}, to
\begin{eqnarray}
  B_0^{\rm sat} \simeq 5.0 \quad \textrm{for} \quad \Lambda_\nu \gg 1,
  \Lambda_\eta < 1 \,.
\label{eq:B0_sat_non_ideal}
\end{eqnarray}


\begin{figure}
  \includegraphics[width=\columnwidth,trim=0 0 0 0]{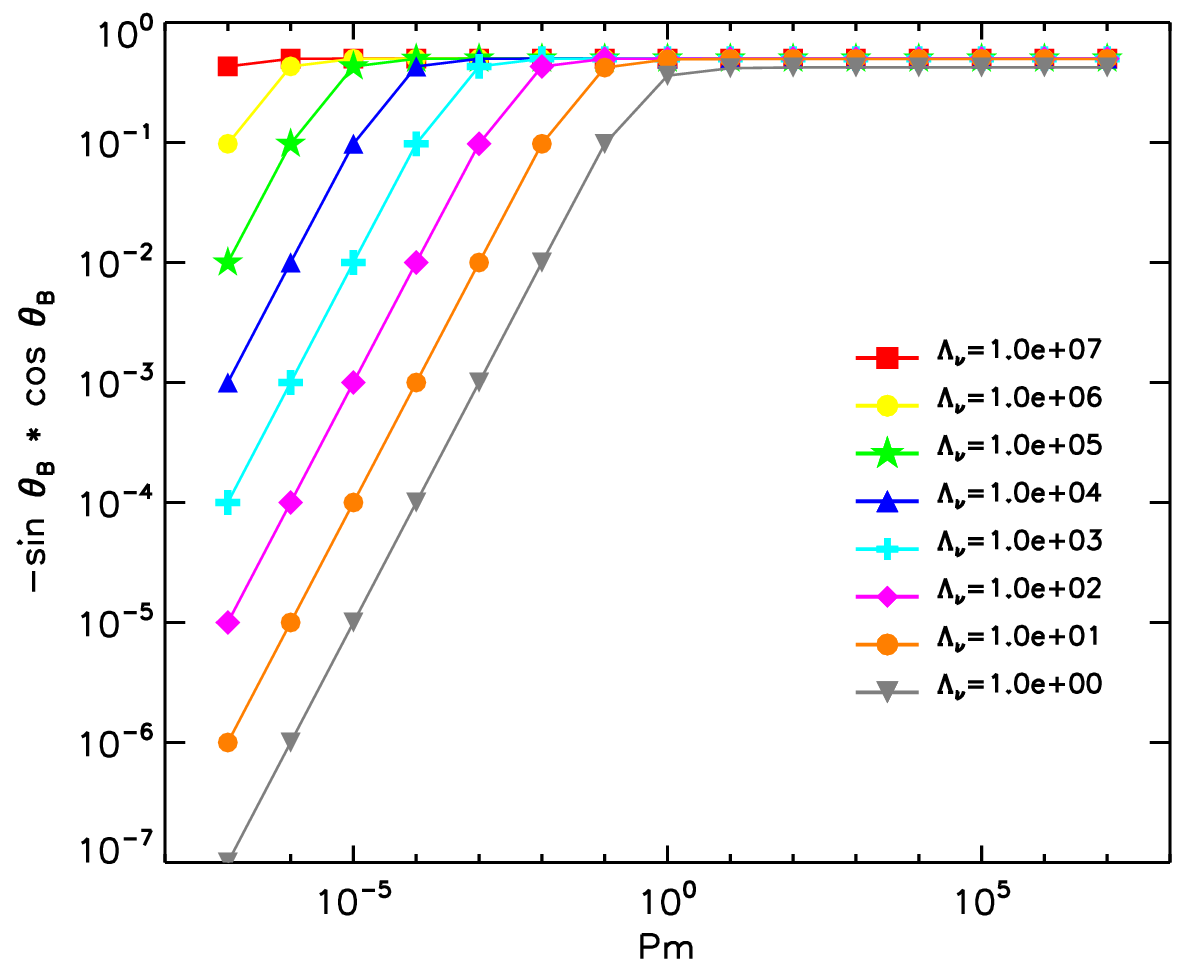}
  \caption{Plot of the function $|\sin\theta_{\rm B} \cos\theta_{\rm
      B}|$, where $\theta_{\rm B}$ corresponds to the angle subtended
    by the magnetic field of the fastest MRI mode and the radial
    direction.  In the ideal limit, with $\Lambda_\eta \gg \Lambda_\nu
    \gg 1$, $|\sin \theta_{\rm B}\cos\theta_{\rm B}| \rightarrow
    1/\sqrt{2}$, while in the inviscid, resistive limit, i.e.,
    $\Lambda_\nu \gg 1$ and $\Lambda_\eta \ll 1$, $|\sin\theta_{\rm
      B}\cos\theta_{\rm B}| \rightarrow {\rm Pm} \, \Lambda_\nu \equiv
    \Lambda_\eta$. Because $B_0^{\rm sat} \rightarrow$ const., the
    behavior of this function is responsible for the decrease of the
    MRI stress shown in Figure~\ref{fig:alpha_beta_sat}.}
  \label{fig:sinOB_cosOB}
\end{figure}

We can formulate a heuristic argument for the existence of an
asymptotic limit for the amplitude $B_0$ in this regime. The analysis
of the classical tearing mode instability, where $B_0 = B_0
\tanh(Kz)$, with $K$ fixed, leads to growth rates $\gamma \sim (\eta
K^2)^{\alpha} (KB_0)^{1-\alpha}$ with $\alpha = 1/2$ or $\alpha=3/5$
depending on the details of how the problem is solved in the resistive
layer (see, e.g., \citealt{FKR63, Sturrock94}).  For $\eta \gg 1 \gg \nu$, there
are only two characteristic scales in the problem, the resistive
timescale $(\eta K^2)^{-1}$ and the Alfv\'en timescale $(KB_0)^{-1}$.
Therefore, the growth rate of the parasites must be $s \sim (\eta
K^2)^{\alpha} (KB_0)^{1-\alpha} = \eta^{\alpha} K^{1+\alpha}
B_0^{1-\alpha} $, with $0\le \alpha\le 1$.  At a fixed scale $K$, the
growth rate of the secondaries is proportional to a positive power of
the resistivity, $s\sim \eta^{\alpha}$.  However, if the scale $K$ is
given by the fastest growing MRI mode then $K \sim \eta^{-1}$ and thus
$s_{\rm max} \sim (B_0)^{1-\alpha}/\eta$.  In this case, because
$\Gamma_{\rm max} \sim \eta^{-1}$, the amplitude $B_0$ at which
$s_{\rm max}=\Gamma_{\rm max}$ must be independent of $\eta$, i.e.,
\begin{eqnarray}
  s_{\rm max} \simeq \Gamma_{\rm max}  \sim
  \left(\eta\, \frac{1}{\eta^2}\right)^\alpha
  \left(\frac{B_0}{\eta}\right)^{1-\alpha} 
  \sim \frac{1}{\eta} \,\,\,.
\end{eqnarray}

Therefore, $B_0 \rightarrow$ const. for $\Lambda_\eta \ll 1$, i.e.,
the amplitude to which the primary MRI magnetic field needs to grow to
in order for both growth rates to become comparable is independent of
the Elsasser number.  This means that, as resistivity increases the
magnetic field that can be generated by the MRI before the tearing
modes become dynamically important reaches an asymptotic, constant
value. However, as we showed in Section~\ref{sec:PI_saturation}, the
associated stress decreases with ${\rm Pm} \, \Lambda_\nu \equiv
\Lambda_\eta$ for $\Lambda_\eta \ll 1$. This is because, in this
limit, $V_0/B_0 \ll 1$ and the stress behaves like $\bar{T}_{r\phi}
\sim B_0^2 |\sin\theta_{\rm B} \cos\theta_{\rm B}| \sim {\rm Pm} \,
\Lambda_\nu \equiv \Lambda_\eta$, see Figure~ \ref{fig:sinOB_cosOB}.

\section{Kelvin-Helmholtz and Tearing Eigenmodes}
\label{sec:eigenmodes}

Throughout this paper we have stated that the fastest parasites are
related to Kelvin-Helmholtz and tearing mode instabilities in the
regimes $\Lambda_\eta \gtrsim 1$ and $\Lambda_\eta < 1$, respectively.
The reason to delay presenting the rigorous evidence supporting these
claims up to this point is based on the existence of the asymptotic
regimes and scaling laws presented in Section~\ref{sec:asymptotia}.
We can now focus on the region of parameter space with ``moderate''
Elsasser number, $\Lambda_\eta \simeq 1$, where changes in
$\Lambda_\eta$ produce non-trivial modifications to the parasitic mode
structure.  With this knowledge, and the insight gained in
Section~\ref{sec:asymptotia}, it is straightforward to describe the
structure of these modes in the limits of large and small Elsasser
numbers.

\subsection{Physical Structure of Parasitic Modes}

In order to understand the nature of the most relevant secondary modes
we analyze the structure of their velocity $\delta \bb{v}_\parallel$
and magnetic fields $\delta \bb{B}_\parallel$ along the directions
associated with their fastest growth, i.e., $\theta = \theta_{\rm
  max}$. For fixed values of the dissipation coefficients, the growth
rates of the secondary instabilities peak around directions which are
almost aligned with either the velocity or magnetic fields of the
primary MRI mode, i.e., $\theta_{\rm max} \simeq \theta_{\rm V}$ for
$\Lambda_\eta \gg 1$ and $\theta_{\rm max} \simeq \theta_{\rm B}$ for
$\Lambda_\eta \ll 1$, see Figure~\ref{fig:s_max_pi}.  Motivated by the
physical characteristics of the Kelvin-Helmholtz and tearing modes, we
calculate the vorticity $\delta \bb{\omega}_\perp$ and current density
$\delta\bb{j}_\perp$ associated with the parasitic modes in the
directions that are perpendicular to the planes defined by $\theta
\equiv \theta_{\rm max}$.  We also analyze the Lagrangian
displacements $\xi_z$ induced by the secondary modes which provide
useful complementary information regarding the reconnection of the
magnetic field associated with the MRI.

Understanding the structure of the secondary modes requires
calculating $\delta \bb{v}(x,y,z)$ and $\delta \bb{b}(x,y,z)$.
However, in the incompressible limit we can calculate $\delta
\bb{v}_\parallel$, $\delta \bb{B}_\parallel$, $\delta
\bb{\omega}_\perp$, and $\delta \bb{j}_\perp$ directly in terms of the
Fourier coefficients for $\delta v_z$ and $\delta b_z$ obtained from
Equations~(\ref{eq:eigensystem_v}) and (\ref{eq:eigensystem_b}).  In
order to do this, it is useful to define a new coordinate system
$(h,p,z)$ that is rotated with respect to $(x,y,z)$ about the
$\check{\bb{z}}$ direction by $\theta_{\rm max}$. We define the
coordinates $h$ and $p$ such that they increase along the direction of
the versors\footnote{In order to simplify notation, in this section we
  refer to the versor associated with the direction of fastest growth,
  i.e.  $\check{\bb{k}}_{\rm h, max}$, simply as $\check{\bb{k}}_{\rm
    h}$} $\check{\bb{k}}_{\rm h}$ and $\check{\bb{k}}_{\rm p}$,
respectively, with $\check{\bb{k}}_{\rm h} \btimes \check{\bb{k}}_{\rm
  p} = \check{\bb{z}}$ and $\check{\bb{k}}_{\rm p} \btimes
\check{\bb{z}} = \check{\bb{k}}_{\rm h}$, see
Figure~\ref{fig:mri_representation_2d}.


\begin{figure*}
  \includegraphics[width=2\columnwidth,trim=0 0 0 0]{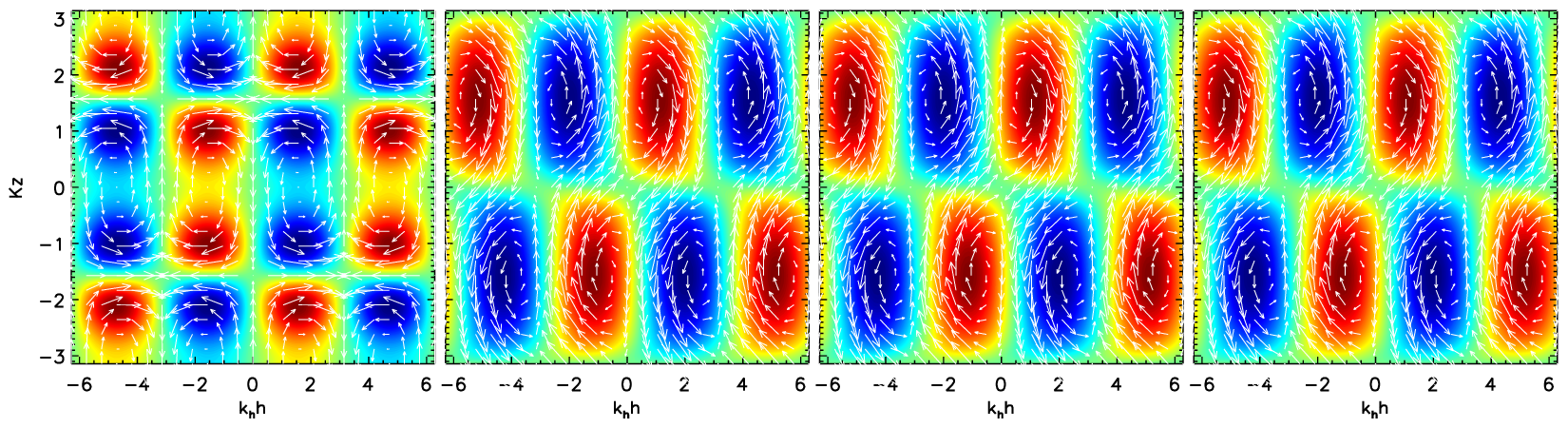}
  \includegraphics[width=2\columnwidth,trim=0 0 0 0]{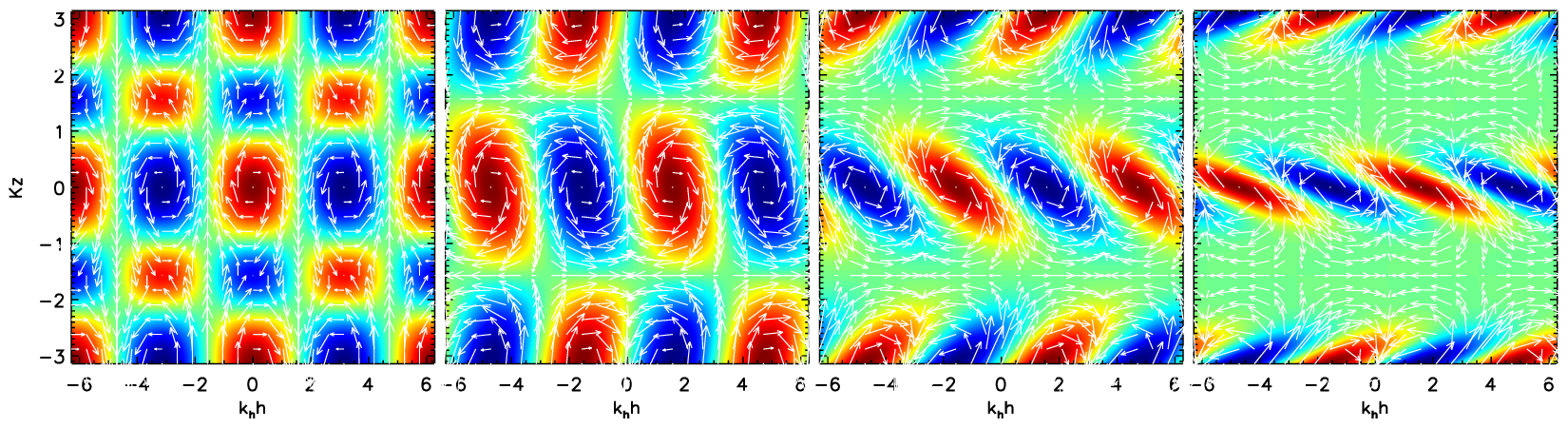}
  \caption{Physical structure of the fastest parasitic modes for
    $\Lambda_\eta = \{0.1, 1, 10, 10^2\}$, from left to right, with
    $\Lambda_\nu \gg 1$.  The arrows in the upper and lower panels
    correspond to the projections of the velocity, $\delta
    \bb{v}_\parallel(h,z)$, and magnetic field, $\delta
    \bb{B}_\parallel(h,z)$, of the parasitic modes onto the plane
    defined by the $z$-axis and the direction $\theta_{\rm max}$.  The
    color contours correspond to the associated vorticity, $\delta
    \bb{\omega}_\perp$, and current density, $\delta \bb{j}_\perp$,
    projected in the direction perpendicular to $\theta_{\rm max}$,
    see Figs.~\ref{fig:mri_representation_3d},
    \ref{fig:mri_representation_2d}, and \ref{fig:s_max_pi}.  The
    left-most (upper and lower) panels, show the vorticity and current
    density patterns characteristic of the tearing mode instability,
    \citep[see Fig.~5.7 in][]{BS03}.  The next set of three upper
    panels show clear signatures of the Kelvin-Helmholtz instability
    in the vorticity contours \citep[see Fig.~1.7.2
    in][]{Batchelor00}.  The lower panels show the (less familiar)
    current density perturbations associated with these
    Kelvin-Helmholtz modes.}
    \label{fig:vorticity_current_no_background}
\end{figure*}

\subsubsection{Velocity and Magnetic Fields}

In the new coordinate system, the velocity and magnetic fields of the
parasitic modes have components
\begin{eqnarray}
\delta \bb{v} &=& 
 \delta v_h \check{\bb{k}}_{\rm h}
+\delta v_p \check{\bb{k}}_{\rm p}
+\delta v_z \check{\bb{z}} \,,\\
\delta \bb{B} &=& 
 \delta B_h \check{\bb{k}}_{\rm h}
+\delta B_p \check{\bb{k}}_{\rm p}
+\delta B_z \check{\bb{z}} \,.
\end{eqnarray}
Since 
$\bb{k}_{\rm h}\bcdot\bb{x} \equiv k_{\rm h} h$, the secondary
perturbations $\delta v_z$ and $\delta B_z$ in
Equations~(\ref{eq:fourier_v}) and (\ref{eq:fourier_b}) become
\begin{eqnarray}
\label{eq:delta_vz_fourier}
\delta v_z(h,z) &=& \sum_{n=-\infty}^{\infty} \alpha_n
e^{-i(n+k_z)z}  e^{-i k_{\rm h} h} \,,\\
\label{eq:delta_Bz_fourier}
\delta B_z(h,z) &=& \sum_{n=-\infty}^{\infty} \beta_n
e^{-i(n+k_z)z}  e^{-i k_{\rm h} h} \,.
\end{eqnarray}
Because we are considering an incompressible fluid, $\delta \bb{v}
\bcdot \check{\bb{k}}_{\rm h} = \partial_z \delta v_z/(ik_{\rm h})$
and $\delta \bb{B} \bcdot \check{\bb{k}}_{\rm h} =
\partial_z \delta B_z/(ik_{\rm h})$. We can thus calculate the
projection of the velocity and magnetic field of a given parasitic
mode onto the plane defined by $(\check{\bb{k}}_{\rm h},
\check{\bb{z}})$, i.e.,
\begin{eqnarray}
\delta \bb{v}_\parallel &=&
(\delta \bb{v} \bcdot \check{\bb{k}}_{\rm h}) \check{\bb{k}}_{\rm h} 
+\delta v_z \check{\bb{z}} \,,\\
\delta \bb{B}_\parallel &=& 
(\delta \bb{B} \bcdot \check{\bb{k}}_{\rm h}) \check{\bb{k}}_{\rm h} 
+\delta B_z \check{\bb{z}} \,,
\end{eqnarray}
just in terms of the Fourier coefficients $\{\alpha_n\}$ and
$\{\beta_n\}$ as
\begin{eqnarray}
\delta \bb{v}_\parallel(h,z) &=& \frac{1}{k_{\rm h}}
\sum_{n=-\infty}^{\infty} (n+k_z) \alpha_n 
e^{-i(n+k_z)z}  e^{-ik_{\rm h} h} \check{\bb{k}}_{\rm h}  \nonumber \\ 
&+&
\sum_{n=-\infty}^{\infty} \alpha_n 
e^{-i(n+k_z)z}  e^{-ik_{\rm h} h} \check{\bb{z}} 
\,,
\label{eq:v_parallel}
\end{eqnarray}
\begin{eqnarray}
\delta \bb{B}_\parallel(h,z) &=& \frac{1}{k_{\rm h}}
\sum_{n=-\infty}^{\infty} (n+k_z) \beta_n 
e^{-i(n+k_z)z}  e^{-ik_{\rm h} h} \check{\bb{k}}_{\rm h} \nonumber \\ 
&+&
\sum_{n=-\infty}^{\infty} \beta_n 
e^{-i(n+k_z)z}  e^{-ik_{\rm h} h} \check{\bb{z}}
\,.
\label{eq:b_parallel}
\end{eqnarray}

Note that the components of the three-dimensional velocity and
magnetic fields that are orthogonal to these planes are independent of
the coordinate $p$. Thus, using the fact that the divergence is
invariant under rotations, we conclude that the divergence of the
two-dimensional vector fields that lie on the plane
$(\check{\bb{k}}_{\rm h}, \check{\bb{z}})$ should vanish, i.e.,
$\del\bcdot\delta\bb{v}_\parallel=\del\bcdot\delta\bb{B}_\parallel=0$.

\subsubsection{Vorticity and Current Density}

The components of the vorticity and current density perpendicular to
the plane defined by $(\check{\bb{k}}_{\rm h}, \check{\bb{z}})$, i.e.,
$\delta \bb{\omega}_\perp = \delta \omega_\perp \check{\bb{k}}_{\rm
  p}$ and $\delta \bb{j}_\perp = \delta j_\perp \check{\bb{k}}_{\rm
  p}$, are given by
\begin{eqnarray}
  \delta \omega_\perp &=& \delta \bb{\omega} \bcdot
  \check{\bb{k}}_{\rm p} = 
  (\del \times \delta \bb{v}) \bcdot \check{\bb{k}}_{\rm p} =
  (\del \times \delta \bb{v}_\parallel) \bcdot \check{\bb{k}}_{\rm p} \,, \\
  \delta j_\perp &=& \delta \bb{j} \bcdot \check{\bb{k}}_{\rm p} = 
  (\del \times \delta \bb{B}) \bcdot \check{\bb{k}}_{\rm p} =
  (\del \times \delta \bb{B}_\parallel) \bcdot \check{\bb{k}}_{\rm p}\,,
\end{eqnarray}
where the action of the curl operators is given by
\begin{eqnarray}
  \del \times \delta \bb{v}_\parallel &=& 
  (\partial_z \delta v_h - \partial_h \delta v_z) \, \check{\bb{k}}_{\rm p} \,, \\
  \del \times \delta \bb{j}_\parallel &=& 
  (\partial_z \delta B_h - \partial_h \delta B_z) \, \check{\bb{k}}_{\rm p} \,.
\end{eqnarray}
Therefore, using Equations~(\ref{eq:v_parallel}) and
(\ref{eq:b_parallel}), we obtain
\begin{eqnarray}
  \delta \bb{\omega}_\perp &=&
  \frac{i \check{\bb{k}}_{\rm p}}{k_{\rm h}}
  \sum_{n=-\infty}^{\infty}  [k_{\rm h}^2+(n+k_z)^2] \alpha_n 
  e^{-i(n+k_z)z}  e^{-ik_{\rm h} h} \,\,\,, \nonumber \\ \\
  \delta \bb{j}_\perp &=&
  \frac{i \check{\bb{k}}_{\rm p}}{k_{\rm h}}
  \sum_{n=-\infty}^{\infty} [k_{\rm h}^2+(n+k_z)^2] \beta_n 
  e^{-i(n+k_z)z}  e^{-ik_{\rm h} h} \,\,\,. \nonumber \\ 
\end{eqnarray}


\begin{figure*}
  \includegraphics[width=2\columnwidth,trim=0 0 0 0]{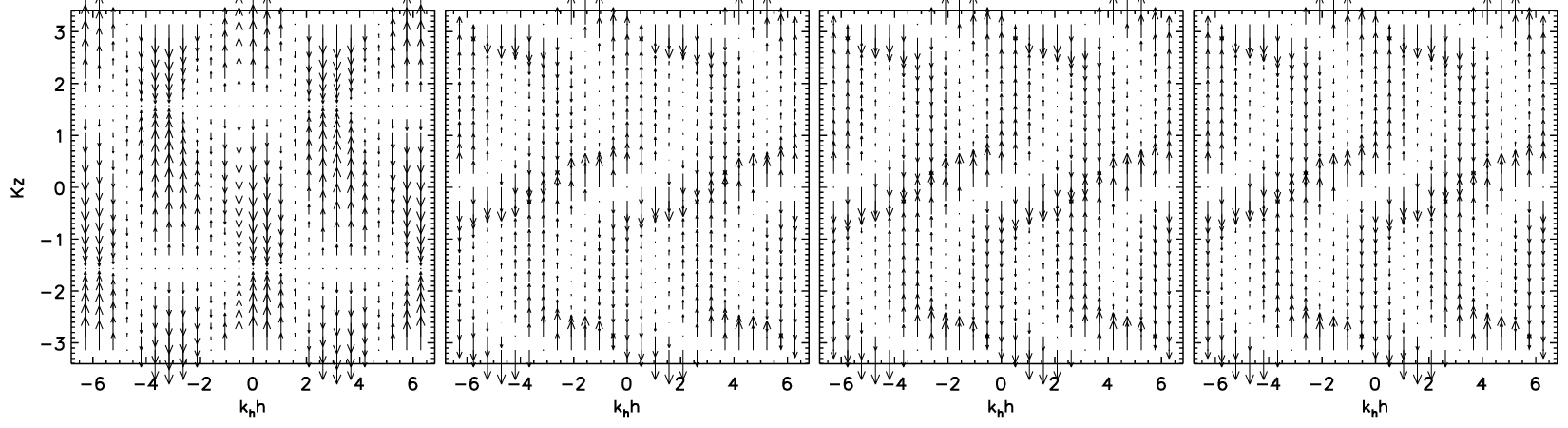}
  \caption{Vertical Lagrangian displacement $\xi_z(h,z)$ corresponding
    to the fastest growing parasitic modes projected onto the plane
    defined by the $z$-axis and the direction $\theta_{\rm max}$. The
    panels correspond to $\Lambda_\eta = \{0.1,1, 10, 10^2\}$,
    from left to right, with $\Lambda_\nu\gg1$.  The convergence of the
    displacement field toward the null surfaces of the MRI magnetic
    field, $\Delta B \propto \cos(Kz)$, located at $K z=\pm n\pi/2$, is
    evident in the leftmost panel. This feature is characteristic of a
    tearing mode; the associated motions are ultimately responsible
    for reconnecting the MRI magnetic field and limiting its growth.
    The next three panels show the displacements associated with the
    fastest Kelvin-Helmholtz modes, which mainly tend to bend the MRI
    magnetic field without directly promoting reconnection.}
\label{fig:lagrangian_z}
\end{figure*}

\subsubsection{Lagrangian Displacement}

The vertical Lagrangian displacement $\xi_z$ provides useful
complementary information to help us identify the physical nature of
the eigenmodes.  The rate of change in the Lagrangian displacement
$\bb{\xi}$ with respect to a point moving with velocity $\bb{v}$ is
given by
\begin{eqnarray}
  \frac{d\bb{\xi}}{dt} \equiv \frac{\partial \bb{\xi}}{\partial t} + (\bb{v}
  \bcdot \del) \bb{\xi} = \delta \bb{v} + \bb{\xi} \bcdot \del \bb{v} \,.
\end{eqnarray}
In the periodic background provided by the primary MRI mode, the
Lagrangian displacement is of the form $\bb{\xi}(\bb{x},t) =
\bb{\xi}_0(z) \, \exp{[s t - i\bb{k}\bcdot\bb{x}]}$ with
$\bb{\xi}_0(\bb{x}+2\pi/K\check{\bb{z}}) = \bb{\xi}_0(\bb{x})$.
Therefore, considering the MRI velocity field, $\Delta \bb{v} =
\bb{V}_0 \sin(Kz)$, as the background velocity, we obtain $\xi_z$ in
terms of the vertical Eulerian velocity $\delta v_z$ in
Equation~(\ref{eq:delta_vz_fourier}) as
\begin{eqnarray}
\label{eq:xi_z}
  \xi_z(h,z) = \frac{\delta v_z(h,z)}{s-i \bb{k}_{\rm h} \bcdot \Delta \bb{v}} \,.
\end{eqnarray}

\subsection{Parasitic Mode Identification}

In Section~\ref{sec:asymptotia} we showed that the eigenvalues
corresponding to the growth rates of the parasitic modes reached well
defined asymptotic regimes for $\Lambda_\eta \ll 1$ and $\Lambda_\eta
\gg 1$ in the limit $\Lambda_\nu \gg 1$. This must also be true for
the eigenmodes. It is then only necessary to explore in detail the
behavior of the modes for moderate values of $\Lambda_\eta$. We thus
focus our attention in the region of parameter space spanned by
$\Lambda_\eta=\{0.1,1,10,10^{2}\}$.  The structures of the fastest
parasitic modes are shown in
Figure~\ref{fig:vorticity_current_no_background}, from left to tight.
The projections of the velocity and magnetic fields onto the planes
defined by $(\check{\bb{k}}_{\rm h}, \check{\bb{z}})$, i.e., $\delta
\bb{v}_\parallel(h,z)$ and $\delta \bb{B}_\parallel(h,z)$, are shown
with white arrows in the upper and lower panels, respectively.  The
color contours show the projection of the vorticity and current
density along the direction $\check{\bb{k}}_{\rm p}$ (perpendicular to
the page), i.e., $\delta \bb{\omega}_\perp$ and $\delta \bb{j}_\perp$.
The red and blue colors correspond to the maximum positive and minimum
negative values associated with the vorticity and current density of
each mode.

\subsubsection{Tearing Modes}

For the Elsasser number $\Lambda_\eta=0.1$, the versor
characterizing the direction of fastest growth, $\check{\bb{k}}_{\rm
  h}$, points in the direction $\theta_{\rm max} \simeq \theta_{\rm
  B}$, see Fig.~(\ref{fig:s_max_pi}) and
Table~\ref{table:mri_pi_mode_data}. This mode, shown in the leftmost
(upper and lower) panels of
Figure~\ref{fig:vorticity_current_no_background}, feeds off the
current density of the primary MRI mode. The corresponding mode
structure resembles closely the perturbations in the current density
and induced vorticity patterns expected in the analysis of the
stability of a set of equidistant current sheets distributed along the
$\check{\bb{z}}$ direction and alternating sense according to $\pm
\check{\bb{k}}_{\rm p}$, \citep[see, Figure 5.7 in][]{BS03}.

The current density of the secondary modes presents maxima and minima
along the planes $Kz=\pm n\pi/2$ where the magnetic field of the
primary mode, $\Delta \bb{B}= \bb{B}_0 \cos(Kz)$, reverses sign.
Thus, the fluctuations induced by these fastest resistive secondary
modes tend to promote reconnection of the MRI field.  This can be
better appreciated by analyzing the vertical Lagrangian displacement
shown in the leftmost panel of Figure~\ref{fig:lagrangian_z}.  There
is another set of periodic maxima and minima in the current density
fluctuations that lie on planes $Kz=0, \pm n\pi$; these are the
locations where the currents associated with the MRI magnetic field
vanish.  Thus, these current density perturbations do not seem to be
due to the unstable configuration presented by the MRI currents
themselves. They rather seem to be needed to satisfy the periodic
constraints on the scale of the unstable MRI mode.

The observed mode structure is qualitatively insensitive to the value
of the Elsasser number as long as $\Lambda_\eta < 1$ and $\Lambda_\nu
\gg 1$, the growth rates and lengthscales associated with each value
of $\Lambda_\eta$ change, of course, as discussed in
Section~\ref{sec:asymptotia}.  We thus conclude that the fastest
parasitic modes correspond to tearing modes for $\Lambda_\eta < 1$.
These parasitic modes are enabled by non-zero resistivity and are thus
absent in the ideal MHD regime studied by \citet{GX94}.


\begin{figure*}
  \includegraphics[width=2\columnwidth,trim=0 0 0 0]{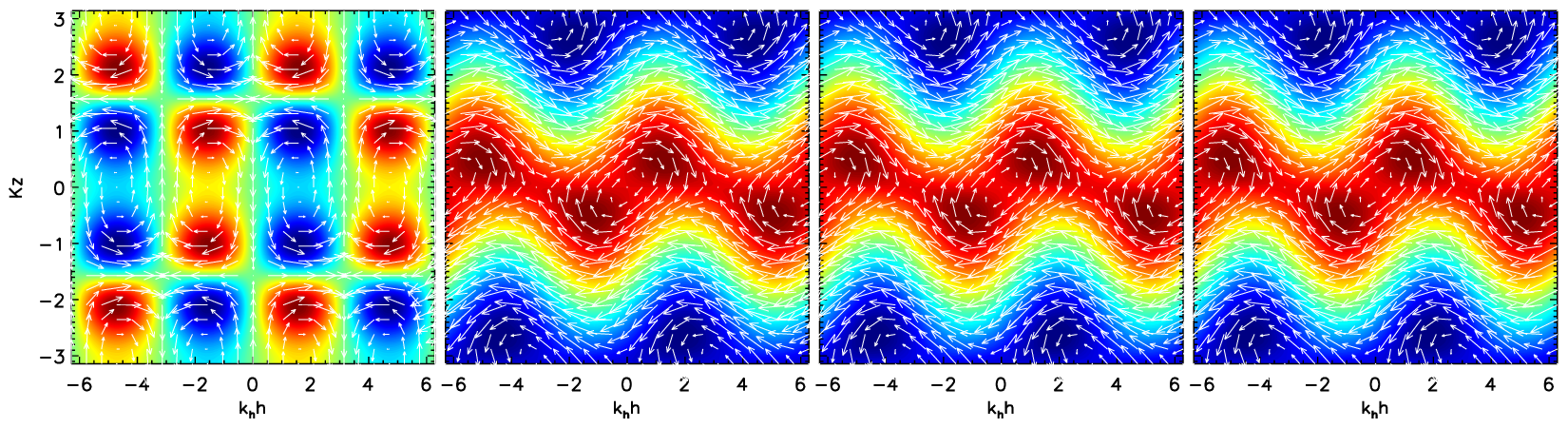}
  \includegraphics[width=2\columnwidth,trim=0 0 0 0]{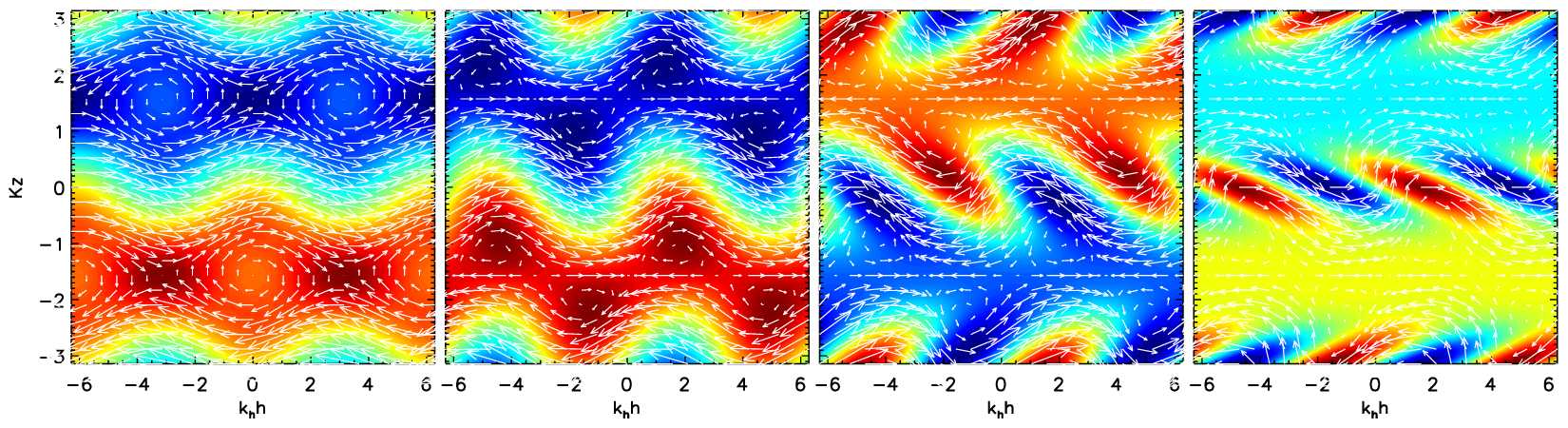}
  \caption{Physical structure of the fastest parasitic modes,
    including the velocity and magnetic fields of the primary MRI
    modes, for $\Lambda_\eta = \{0.1, 1, 10, 10^2\}$, from left to
    right, with $\Lambda_\nu\gg1$.  The arrows in the upper and lower
    panels correspond, respectively, to the projections of the total
    (primary plus secondary) velocity, $\Delta\bb{V}_{\parallel} +
    \delta \bb{v}_\parallel(h,z)$, and magnetic fields,
    $\Delta\bb{B}_{\parallel} + \delta \bb{B}_\parallel(h,z)$, onto
    the plane defined by the $z$-axis and the direction $\theta_{\rm
      max}$. The color contours correspond to the associated total
    vorticity and total current density projected onto the direction
    perpendicular to $\theta_{\rm max}$.}
  \label{fig:vorticity_current_background}
\end{figure*}

\subsubsection{Kelvin-Helmholtz Modes}

For the Elsasser numbers $\Lambda_\eta=\{1,10,10^2\}$, the versors
$\check{\bb{k}}_{\rm h}$ characterizing the direction of fastest
growth point in the direction $\theta_{\rm max} \simeq \theta_{\rm V}$
for the three rightmost sets of panels in Fig.~(\ref{fig:s_max_pi}),
see also Table~\ref{table:mri_pi_mode_data}.  These modes, shown in
the three rightmost (upper and lower) panels of
Figure~\ref{fig:vorticity_current_no_background}, feed off the shear
in the velocity field of the corresponding primary MRI modes.  The
velocity and vorticity fields show a periodic structure similar to
what is expected from the stability analysis of a periodic set of
equidistant vortex sheets distributed along the $\check{\bb{z}}$
direction and alternating sense according to $\pm \check{\bb{k}}_{\rm
  p}$.

The mode structure in the velocity and vorticity fields, as well as
the growth rates and lengthscales, associated with each value of
$\Lambda_\eta$ are quantitatively insensitive to the value of the
Elsasser number as long as $\Lambda_\eta \ge 1$ and $\Lambda_\nu \gg
1$, as discussed in Section~\ref{sec:asymptotia}.  The structure of
the current densities associated with these modes show some evolution
as a function of $\Lambda_\eta > 1$.  However, these modes are not
intrinsically modified from pure Kelvin-Helmholtz modes since the
amplitude of the fluctuations in the magnetic and current density
fields is much smaller than the fluctuations in the velocity and
vorticity fields.  We thus conclude that the fastest parasitic modes
correspond to Kelvin-Helmholtz modes for $\Lambda_\eta > 1$. In the
limit $\Lambda_\eta \gg 1$, these correspond of course to the
Kelvin-Helmholtz modes alluded to in \citet{GX94}.

The current density of the secondary modes vanishes along, and also in
the vicinity of, the planes $z=\pm n\pi/2$ where the magnetic field of
the primary mode, $\Delta \bb{B}= \bb{B}_0 \cos(Kz)$, changes sign.
The corresponding Lagrangian displacement associated with these modes
for $\Lambda_\eta=\{1,10,10^2\}$ are shown in the three rightmost
panels in Figure~\ref{fig:lagrangian_z}, respectively. The
fluctuations induced by the fastest growing Kelvin-Helmholtz secondary
instabilities tend to bend the horizontal MRI magnetic field without
directly promoting their reconnection.

\subsubsection{Inclusion of MRI-background Fields}

The white arrows in the upper and lower panels of
Figure~\ref{fig:vorticity_current_background} show the structure of
the velocity and magnetic fields in
Figure~\ref{fig:vorticity_current_no_background} when the background
MRI fields are added to the secondary modes.  The projections of the
total velocity and magnetic fields onto the planes defined by
$(\check{\bb{k}}_{\rm h}, \check{\bb{z}})$, are given by
$\Delta\bb{V}_{\parallel} + \delta \bb{v}_\parallel(h,z)$ and
$\Delta\bb{B}_{\parallel} + \delta \bb{B}_\parallel(h,z)$, where
\begin{eqnarray}
\label{eq:V0_parallel}
  \Delta\bb{V}_{\parallel} &=& \bb{V}_0\sin(Kz) \cos(\theta_{\rm max}-\theta_{\rm V}) \,, \\
\label{eq:B0_parallel}
  \Delta\bb{B}_{\parallel} &=& \bb{B}_0\cos(Kz) \cos(\theta_{\rm max}-\theta_{\rm B}) \,. 
\end{eqnarray}
The color contours show the projection of the total vorticity and
current density along the direction $\check{\bb{k}}_{\rm p}$
(perpendicular to the page), i.e., $\delta \bb{\omega}_{\perp, 0} +
\delta \bb{\omega}_\perp$ and $\delta \bb{j}_{\perp, 0} + \delta
\bb{j}_\perp$. The contributions of the primary MRI mode are given by
\begin{eqnarray}
\label{eq:omega0_perp}
\delta \bb{\omega}_{0,\perp} = \del \times \Delta\bb{V}_{\parallel} &=& 
V_0 \cos(Kz) \cos(\theta - \theta_{\rm V}) \, \check{\bb{k}}_{\rm p}  \,, \\
\label{eq:j0_perp}
\delta \bb{j}_{0,\perp} = \del \times \Delta\bb{B}_{\parallel} &=& 
- B_0 \sin(Kz) \cos(\theta - \theta_{\rm B}) \, \check{\bb{k}}_{\rm p} \,.
\end{eqnarray}
In all cases, the amplitude of the primary mode $V_0$ and $B_0$ is
such that the fastest secondary modes grow as fast as the primary MRI
mode, i.e., $s_{\rm max}(\nu,\eta, K_{\rm max}) = \Gamma_{\rm max}
(\nu,\eta)$.

The addition of the MRI fields facilitates the identification of the
velocity and magnetic fields that result from the influence of the
secondary instabilities with the more familiar structures that are
expected from Kelvin-Helmholtz and tearing mode instabilities in
periodic backgrounds. In particular, it highlights the presence of
``$O$'' and ``$X$'' points in the case of the tearing mode and the
wave-like structure of the velocity field in the cases associated with
Kelvin-Helmholtz modes.

\section{Discussion}
\label{sec:discussion}

\subsection{This Work in Context}

Understanding from first principles the saturation of MHD turbulence,
and the associated angular momentum transport, driven by the MRI in
astrophysical disks surrounding young stars and compact objects is a
challenging endeavor. The large dynamical range in temperatures,
densities, and magnetic fields in these accretion disks implies that
the dimensionless variables parameterizing dissipation span a wide
range of values.  Numerical simulations are indispensable tools for
understanding the properties of MRI-driven turbulence in different
environments.  However, most numerical studies to date have been
carried out either without physical dissipation or with dissipation
coefficients that have a small overlap with the regions of parameter
space relevant to either astrophysical or experimental setups, see
Section~\ref{sec:parameter_space}.  In order to gain physical insight
into the regimes of interest, it seems necessary to combine the
results of numerical simulations with analytical efforts aim at
identifying the essential processes at work.

The idea that secondary instabilities can limit the growth of the MRI,
and (more speculatively) play a role in the subsequent turbulent
state, was put forward by \citet{GX94} and has been considered in more
detail by a number of works more recently (\citealt{Sano07,Bodoetal08,
  Obergaulingeretal09, LLB09, PG09}; c.f., \citealt{KJ05, JJK08a,
  JJK08b}).  In this paper, we carried out a thorough study of the
spectrum and physical structure of parasitic modes that feed off exact
MRI modes when the amplitude of the magnetic field produced by the MRI
is such that the instantaneous growth rate of the fastest parasitic
mode matches that of the fastest primary mode.  Following
\citet{PG09}, we argued that this ``saturation'' amplitude provides an
estimate of the magnetic field that can be generated by the MRI before
the secondary instabilities suppress its growth significantly.  While
we invoked several assumptions and approximations in order to make the
problem tractable, our approach enabled us to explore dissipative
regimes that are relevant to astrophysical and laboratory conditions
that lie beyond the regime accessible to current numerical
simulations.

We mention two limitations imposed by our assumptions that are worth
emphasizing due to their potential significance (see also the related
discussion in Section 2.4 in \citealt{LLB09}): (i) We neglected the
effects of shear on the dynamics of the secondary modes.
Non-axisymmetric parasitic modes will shear linearly in time
\citep{GX94}; therefore, assuming fixed horizontal versors for the
parasites is an approximation.  (ii) We ignored the explicit coupling
between the evolution of the MRI-modes and the secondary
instabilities. As the secondary modes grow they drain energy from the
primary modes; therefore their growth rates, which rely on the
amplitude of the MRI, could be affected. This approximation might also
affect the estimates of the saturation amplitude of the MRI modes,
since they provide the source of energy that feeds the parasites.
Thus the extrapolation of the results presented here to the non-linear
regime should be complemented with the pertinent quota of skepticism.
Having said this, at present, the properties of parasitic modes
described here provide valuable analytical guidance and a basic
framework to design and interpret tailored numerical experiments in
order to shed light into the non-linear saturation of the MRI.

We summarize here our results and explain how several features of
numerical simulations designed to address the saturation of the MRI in
protoplanetary disks and accretion disks surrounding compact objects
can be interpreted in terms of our findings.

\subsection{Results, Scaling Laws, and Parasitic Mode Identification}

When the magnetic fields involved are weak enough so that the
incompressible limit holds, the dynamics of the MRI and the parasitic
instabilities depend only on any two independent dimensionless numbers
that can be formed using the Elsasser number, $\Lambda_\eta \equiv
\bar{v}_{{\rm A}z}^2/(\eta\Omega_0)$, and is its viscous counterpart,
$\Lambda_\nu\equiv \bar{v}_{{\rm A}z}^2/(\nu\Omega_0)$.  Motivated by
recent works that suggest that the saturation amplitude of the MRI
depends on the magnetic Prandtl number \citep{UMR07,LL07}, we
considered $\Lambda_\nu$ and ${\rm Pm} \equiv
\Lambda_\eta/\Lambda_\nu$ as independent parameters.  

\begin{figure}
  \includegraphics[width=\columnwidth,trim=0 0 0 0]{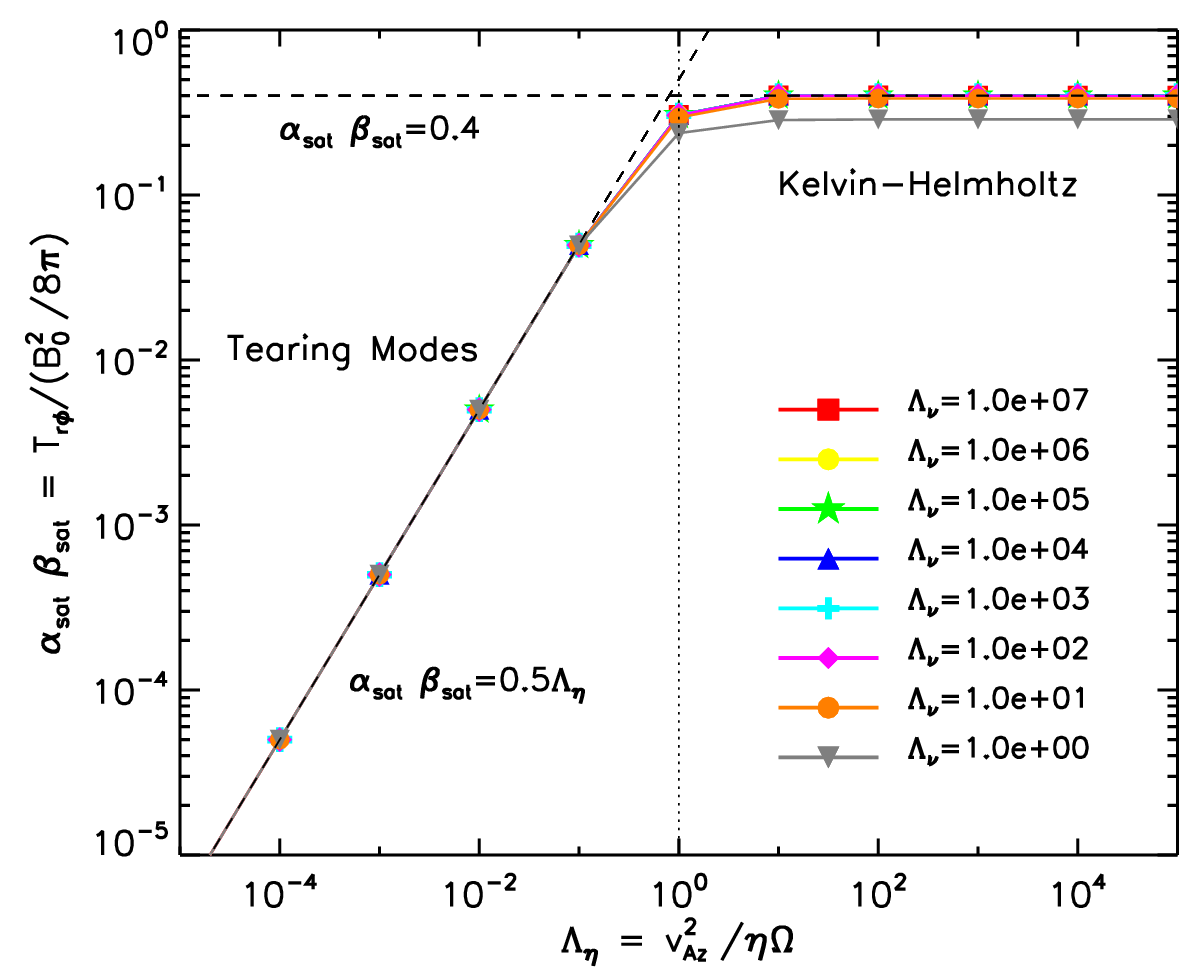}
  \caption{Predicted values for the ratio of stress to magnetic energy
    density, i.e., the product $\alpha_{\rm sat}\beta_{\rm sat}$, if
    saturation occurs when the fastest parasitic and primary MRI
    growth rates match.  For $\Lambda_\nu \gtrsim 10$,
    Equations~(\ref{eq:alpha_beta_sat_ideal_elsasser}) and
    (\ref{eq:alpha_beta_sat_resis_elsasser}) describe the results of
    the parasitic mode analysis remarkably well.  In the quasi-ideal
    MHD limit, applicable to the fully ionized regions of accretion
    disks, $\alpha_{\rm sat}\beta_{\rm sat} = 0.4$. In the inviscid,
    resistive limit, applicable to poorly ionized regions of
    protoplanetary disks $\alpha_{\rm sat}\beta_{\rm sat} = 0.5 \,
    \Lambda_\eta$.  These results are consistent with the numerical
    simulations shown in Figure~20 of \citet{SS02}, see the discussion
    in the text for more details. The modes responsible for saturation
    correspond to Kelvin-Helmholtz and tearing modes for Elsasser
    numbers $\Lambda_\eta$ larger and smaller than unity,
    respectively. The Elsasser number for current MRI experiments is
    close to unity. In this regime, both types of modes present
    similar growth rates.}
  \label{fig:alpha_beta_sat_elsasser}
\vspace{2.5mm}
\end{figure}

We found, however, that the parameter driving the behavior of the
growth rates of the parasites and the MRI, and thus the magnetic
energy density and stresses at saturation, is the Elsasser number
$\Lambda_\eta$.  In particular, we found that, as long as viscous
dissipation is small, i.e., $\Lambda_\nu \gtrsim 10$, then there
exists two regimes (see Figure~\ref{fig:alpha_beta_sat_elsasser}): (i)
quasi-ideal MHD, where the physical properties of the MRI and
parasitic instabilities are insensitive to dissipation. This holds as
long as $\Lambda_\eta > 1$, which is applicable to the fully
ionized regions of accretion disks around compact objects. (ii)
inviscid, resistive MHD, where all the relevant dependencies on
$\Lambda_\nu$ and ${\rm Pm}$ are only through the product ${\rm Pm} \,
\Lambda_\nu$, i.e., the Elsasser number $\Lambda_\eta$.  This regime
corresponds to $\Lambda_\eta < 1$, and characterizes poorly ionized
regions of protoplanetary disks \citep{Jin96, SM99}. The Elsasser
number for current Taylor-Couette MRI experiments is close to unity
\citep{Nornbergetal10}. In this regime, both types of modes present
similar growth rates.

We estimated the amplitude of the magnetic fields and stresses
generated by the MRI when the secondary instabilities become
dynamically important.  The saturation amplitude of the magnetic
fields is fairly insensitive to dissipation with $B_0^{\rm
  sat}/\bar{B}_z \simeq 4$--$5$.  However, the stress behaves very
differently.  For $\Lambda_\eta > 1$ the stress reaches and asymptotic
value independent of the dissipation coefficients, while for
$\Lambda_\eta < 1$ the stress decreases linearly with decreasing
Elsasser number, i.e., $\bar{T}^{\rm sat}_{r\phi} \propto
\Lambda_\eta$. This result is consistent with the numerical
simulations of resistive MHD shearing boxes carried out by
\citet{FSH00} and \citet{SS02} discussed below.  We calculated the
ratio between the stress and the magnetic energy density associated
with the primary MRI mode in two regimes of interest and obtained
\begin{eqnarray}
\label{eq:alpha_beta_sat_ideal_elsasser}
  \alpha_{\rm sat}\beta_{\rm sat} &\simeq& 0.4 \,\,\quad \quad  \textrm{for} \quad \Lambda_\nu \gg 1 \,, \Lambda_\eta >1 \,, \\
\label{eq:alpha_beta_sat_resis_elsasser}
  \alpha_{\rm sat}\beta_{\rm sat} &\simeq& 0.5 \, \Lambda_\eta \quad \textrm{for} \quad \Lambda_\nu \gg 1 \,, \Lambda_\eta<1 \,.
\end{eqnarray}
This anti-correlation between $\alpha_{\rm sat}$ and $\beta_{\rm sat}$
is seen in the numerical simulations of 'ideal' MHD (i.e., without
explicit dissipation) in \citet{HGB95}, as well as in the resistive
runs in \citet{SIM98}.  We understood these results in terms of the
behavior of both primary and secondary modes.

We showed that important properties of the fastest secondary
instabilities found in \citet{PG09} for the region of parameter space
accessible to current numerical simulations are generic.  The fastest
parasitic modes are non-axisymmetric, have purely real growth rates,
have the same vertical periodicity as the primary MRI mode, and
horizontal wavelengths that are roughly twice as long.  Their
wavevectors $\bb{k}_{\rm h}$ are nearly aligned with either the
velocity or the magnetic field of the primary mode.  The first type
dominate for $\Lambda_\eta \ge 1$ and correspond to Kelvin-Helmholtz
modes that feed off the velocity shear induced by the MRI and reach
their maximum growth rates along the direction of the MRI velocity
field.  The second type dominate for $\Lambda_\eta < 1$ and are
related to tearing modes that feed off the MRI currents and grow the
fastest along the direction of the MRI magnetic field.

We emphasized the importance of understanding the structure of the
vorticity and current density patterns associated with the secondary
instabilities, as they provide a mean to confirm their association
with Kelvin-Helmholtz and tearing mode instabilities.  The analysis
presented in Section~\ref{sec:eigenmodes} suggests a strategy to
recognize the presence of parasitic modes and attempt to identify
their nature: (i) Evolve the simulation until the breakdown of the
initial exponential growth or subsequent peaks in stress or magnetic
energy. (ii) Dissect the simulation domain in planes perpendicular to
the mid-plane that contain the $z$-axis.  (iii) Project the velocity
and magnetic field in these planes and take the curls in order to
obtain the corresponding vorticity and current density. (iv) Determine
whether these resemble what is expected from the Kelvin-Helmholtz or
tearing mode instabilities.

\subsection{Connection to Previous Works}

\subsubsection{Evolution of MRI in Two and Three Dimensions, Extent of
Simulation Domains, and Aspect Ratios}

Exploiting the numerical advantages of working with two-dimensional,
axisymmetric simulations, \citet{MS08} explored the effects of
explicit viscosity with $\Lambda_\nu = \{0.01, 0.1, 1, 10\}$ (and no
explicit resistivity) and explicit resistivity with $\Lambda_\eta =
\{0.1, 0.3, 1, 10, 100\}$ (and no explicit viscosity).  The fact that
the fastest secondary modes are not axisymmetric suggests that the
transition to turbulence, and perhaps the subsequent non-linear
evolution, should be rather different in two-dimensional, axisymmetric
and three-dimensional numerical simulations, see e.g., \citet{GG08,
  Obergaulingeretal09}.  Indeed, the viscous simulations shown in
Figures 4a and 7 in \citet{MS08} do not seem to reach saturation.

The wavelength of the fastest growing MRI mode increases as viscosity
and resistivity increase \citep{PC08}.  If the vertical extent of the
simulation domain is not large enough, the fastest growing mode, or
even the smallest unstable mode, might not fit in the domain. This
might lead to spurious dependencies of the saturation on the
dissipation coefficients (see below). Two- and three-dimensional
simulations with aspect ratios $L_r/L_z = 1$ seem to evolve
differently than those with $L_r/L_z \simeq 2$. This has been seen in
both types of setups by \citet{Obergaulingeretal09} and
\citet{Bodoetal08}. This could be due the fact that both axisymmetric
and the fastest growing (non-axisymmetric) parasitic modes have
horizontal wavelengths larger than the vertical wavelength of the
primary MRI mode. However, once the fastest parasites are allowed to
evolve unimpeded, increasing the aspect ratio further should not alter
the results significantly.

Therefore, in order to reliably assess the dependence of the
saturation of the MRI on the value of the dissipation coefficients,
three-dimensional domains with $L_z > \lambda_{\rm MRI} \equiv
2\pi/K_{\rm max}(\nu,\eta)$ and $L_r/L_z>2$ seem to be required.

\subsubsection{Transition to Turbulence and Recurrence of Channels}

As resistivity increases, the timescale for the MRI to grow to
amplitudes such that the parasites become dynamically important also
increases. This behavior is in agreement with the results presented in
\citet{FSH00}.  Their Figure~2 shows that the mechanism that disrupts
the initial exponential growth of the MRI (arguably related to the
instabilities under study) is sensitive to the values of microphysical
dissipation, with more resistive runs reaching higher amplitudes.  

The ratio of the peak magnetic energy density between their more
resistive runs, corresponding to\footnote{Notice that their definition
  of magnetic Reynolds number ${\rm Rm}$ is related to our definition
  of Elsasser number via $\Lambda_\eta \equiv 2{\rm Rm}/\beta$ with
  $\beta=400$.} $\Lambda_\eta \simeq 1$, and the least resistive
simulation, which behaves similarly to their simulation without
explicit dissipation, is about ten.  This is in contrast to our
expectation that these saturation amplitudes should not differ by more
than a factor of two, see Figure~\ref{fig:B0_sat}.  Note however, that
the fastest growing mode in their 'ideal' run corresponds to
$\lambda_{\rm MRI} = L_z/2$ while in the very resistive runs
$\lambda_{\rm MRI} = L_z$. This means that the fastest parasites,
which quite generically posses horizontal wavelengths that are roughly
a factor of two longer than the fastest growing MRI mode, might be
suppressed in the more resistive simulations, which could account for
the larger amplitudes observed.

\citet{FSH00} and \citet{SI01} studied the effects of resistivity in
the subsequent recurrent emergence of organized fluid motions with
highly correlated magnetic fields (so called "channels"). The
resistive simulations of \citet{FSH00} show the same type of
quasi-periodic variations in the mean stress observed in 'ideal' MHD
simulations after the break up of the initial exponential growth.  The
amplitude of the fluctuations around the mean and the timescales
involved increase as resistivity increases, see their Figure~2. This
type of fluctuations are known to be reduced significantly when the
aspect ratio $L_r/L_z$ increases beyond a factor of two
\citep{Bodoetal08}.  For the more resistive simulations, the
wavelength of the fastest MRI mode is $\lambda_{\rm MRI} = L_z$.  It
is thus tempting to attribute the persistence of these fluctuations to
the suppression of the fastest parasitic modes in the simulations with
$L_r/L_z=1$.

The parasitic instabilities limit the amount of energy that the MRI
can extract from the the differential rotation.  This suggests that
the larger peak amplitudes of the fluctuations in the more resistive
runs could be due to the fact that the timescale for the MRI to grow
to amplitudes such that the parasites become dynamically important
increases with resistivity.  Once the ordered motions of the primary
modes are disrupted by the fastest available parasitic modes,
non-linear interactions are arguably responsible for the cascade of
energy to smaller scales where it dissipates.  Since more magnetic
energy needs to be dissipated after the reconnection of recurrent
fluctuations with higher amplitudes, the longer timescales involved in
the more resistive runs seems natural.

The fact that the peak amplitude of the initial exponential growth is
larger than the subsequent fluctuations can be explained in terms of
the corresponding amplitude of the seed fluctuations that excite the
parasites. The amplitude to which the MRI can grow before the
parasites reach similar amplitudes is sensitive to the initial
amplitude of the seed fluctuations that excite the secondary
instabilities. In the absence of explicit perturbations to seed the
secondary instabilities in the linear regime of the MRI, the amplitude
of the fastest growing MRI mode could overshoot our estimates by a
large factor, which depends logarithmically on the amplitude of the
seed fluctuations \citep{PG09}.  The subsequent channels are emerging
from a turbulent background where the seed fluctuations for the
parasites is larger.  Therefore these parasites will be able to reach
an amplitude similar to that of the dominant primary MRI mode faster
than the parasites responsible for halting the initial exponential
growth.  This would result in smaller amplitudes for the subsequent
channels (c.f.  \citet{LLB09}, where it is suggested that the smaller
amplitude of the recurrent channels is due to interactions between
modes.)

There is numerical evidence that suggests that ohmic heating due to
the reconnection of MRI field lines is an important source of energy
in resistive MHD \citep{FSH00,SI01}.  Although this reconnection
process has been attributed to the ideal parasitic instabilities
studied in \citet{GX94}, this is a clear signature of non-ideal MHD
effects.  In allowing for non-zero resistivity and calculating the
currents associated with the secondary instabilities, we have gone one
step forward in establishing the chain of processes that enable the
conversion of gravitational energy into thermal energy in
differentially rotating, magnetized, non-ideal plasmas. The properties
of tearing modes discussed in Section~\ref{sec:eigenmodes} should
provide better guidance for interpreting these reconnection events.

\subsubsection{Dependence of Saturation on Dissipation Coefficients:
  Elsasser Number vs. Magnetic Prandtl Number}

In the region of parameter space where our analysis overlaps with the
regime accessible to the numerical simulations of \citet{LL07}, the
values of $\alpha_{\rm sat}$ are within factors of a few of the values
obtained in the turbulent regime. The dependence on the stresses at
saturation on the magnetic Prandtl number in our calculations is less
pronounced than what they report and the predicted value of
$\alpha_{\rm sat}$ is smaller, by a factor of 6 at ${\rm Pm} =1$. Part
of these differences can be accounted for with a more sensible
operational definition of saturation. \citet{PG09} provide a way to
estimate the amplitude of the fields at saturation when primary and
secondary modes reach comparable amplitudes based on the values
obtained when they reach equal growth rates. They conclude that the
overshoot factor, which depends logarithmically on the initial
amplitude of the parasite, is likely between $\simeq 3$ and $\simeq
10$.  These arguments are applicable quite generically since they
mainly rely on the fact that the growth rate of the fastest secondary
modes is linear in the amplitude of the primary MRI mode, see
Section~\ref{sec:asymptotia}.

We posit that it is then conceivable that
Equations~(\ref{eq:alpha_beta_sat_ideal_elsasser}) and
(\ref{eq:alpha_beta_sat_resis_elsasser}) could provide, within factors
of a few, a reasonable description of the saturation level of the MRI
in a wide region of parameter space. Our results suggests that, as
long as viscous dissipation does not dominate the dynamics of the
fluid, i.e.  $\Lambda_\nu \ge 10$, which is the case in many
astrophysical environments, as well as in laboratory experiments, see
Section~\ref{sec:PI_saturation}, then the angular momentum transport
due to the MRI depends mainly on the Elsasser number, see
Figure~\ref{fig:alpha_beta_sat_elsasser}.  The numerical results
presented in \citet{SS02} and \citet{FSH00} seem to support this
statement.

\citet{SS02} carried out an extensive numerical study of the
saturation of the MRI considering ohmic dissipation and Hall terms,
with no explicit viscosity.  They identify the existence of a critical
Elsasser number of order unity, $\Lambda_\eta^{\rm c} \simeq 1$, which
is independent of the strength and geometry of the magnetic field or
the magnitude of the Hall term.  For Elsasser numbers $\Lambda_\eta >
1$, the stress at saturation is rather insensitive to $\Lambda_\eta$.
However, for $\Lambda_\eta < 1$, they find that the mean value of the
stress decreases linearly with the Elsasser number, i.e.,
$\bar{T}^{\rm sat}_{r\phi} \propto \Lambda_\eta$, see their Figure~20.
This result describes numerical simulations spanning a wide range of
Elsasser numbers and magnetic field configurations\footnote{Notice
  that the subset of simulations with vertical background fields in
  \citet{SS02}, show a dependence with $\Lambda_\eta$ that is steeper
  than linear.  The inspection of their Table~1 suggests that this
  could be due to: (i) resolution constrains, for the simulations
  where several fast growing MRI modes fit within the box, (ii) aspect
  ratios constrains, for the simulations where the most unstable mode
  is well resolved but its vertical wavelength is of the order of
  $L_z$, or (iii) because the most unstable mode does not fit within
  the box.}. The inclusion of Hall terms does not seem to affect these
results significantly.  Even though the most resistive simulations
carried out in \citet{FSH00} correspond to Elsasser numbers slightly
larger than unity, their Table~1 shows that the stress at saturation
decreases linearly with increasing Elsasser number.

Recent works investigating the effects of dissipation on the
saturation of the MRI suggest that the saturation amplitude depends on
viscosity and resistivity mainly through the magnetic Prandtl number,
${\rm Pm} \equiv \Lambda_\eta/\Lambda_\nu$.  \citet{UMR07} performed a
weakly non-linear analysis of the viscous, resistive MRI in the limit
of small magnetic Prandtl number, i.e., ${\rm Pm} \ll 1$. They found
that the saturation amplitude is proportional to ${\rm Pm}^{1/2}$,
while the associated momentum transport scales as ${\rm Re}^{-1}$,
where ${\rm Re}\equiv \Omega_0 L^2/\nu$ is the Reynolds number.
\citet{LL07} carried out a series of shearing box simulations in
incompressible MHD with explicit dissipation. In the range of
parameters that they explored, i.e., $1 \lesssim \Lambda_\nu \lesssim
100$ and $0.1 \lesssim {\rm Pm} \lesssim 10$, the stress decreases
with decreasing magnetic Prandtl number as ${\rm Pm}^\delta$ with
$\delta\simeq 0.25$--$0.5$, with a weak dependence on the Reynolds
number. Although it must be noted that the two approaches (weakly non-linear
analysis vs. three-dimensional simulations) are indeed quite
different, \citet{LL07} suggest that the difference between their
findings and those of \citet{UMR07} could be related to boundary
conditions.

It seems reasonable to argue that the saturation of the MHD turbulence
driven by the MRI at large values of magnetic Prandtl number with
$\Lambda_\eta > \Lambda_\nu \gg 1$ should converge towards an
asymptotic value.  This limit does not seem to have been achieved in
Figure~10 in \citet{LL07} . Perhaps the highest Reynolds and magnetic
Prandtl numbers considered, which correspond in our definitions to
$\Lambda_\nu \simeq 10^2$ and ${\rm Pm} \simeq 10$, are not large
enough to observe the asymptotic limit suggested by our analysis (see
also their discussion regarding the effects of limited resolution for
this case).  

It is tempting to examine the seemingly discrepant dependencies of
saturation on Elsasser and magnetic Prandtl number for the most
dissipative simulations in \citet{FSH00} and \citet{LL07}.
\citet{FSH00} considered ohmic dissipation but not explicit viscosity
and it is thus hard to assign a well defined magnetic Prandtl number
to their simulations.  However, simulations without explicit
dissipation seem to be characterized by an effective magnetic Prandtl
number of order unity \citep{SHB09}.  We could speculate that the
simulations with explicit resistivity but no explicit viscosity in
\citet{FSH00} correspond to $\Lambda_\eta < \Lambda_\nu$, and thus the
regime $\Lambda_\eta \simeq 1$ corresponds to ${\rm Pm} \lesssim 1$.
However, the most inviscid and resistive simulations in \citet{LL07}
correspond to $\Lambda_\nu \simeq 10^2$ and ${\rm Pm} \simeq 0.1$,
i.e., $\Lambda_\eta \simeq 10$.  Thus, even though the viscosity seems
small in these simulations, the Elsasser number does not seem to be
low enough for the stress to show the behavior $\bar{T}^{\rm
  sat}_{r\phi} \propto \Lambda_\eta$ for $\Lambda_\eta < 1$ found in
\citet{SS02}.

Our results suggest that the Elsasser number dictates the saturation
level of the angular momentum transport driven by MHD turbulence in
astrophysical disks and experiments, see
Figure~\ref{fig:alpha_beta_sat_elsasser}.  Despite the inherent
limitations of the parasitic mode analysis presented here, this result
seems to be supported by numerical simulations of \citet{FSH00} and
\citet{SS02}.  However, given the rather small dynamical range of
dissipation coefficients that can be currently explored, these
conclusions warrant further examination.
Additional numerical studies with explicit viscosity and resistivity
in three-dimensional domains (that can accommodate for the most
relevant MRI and parasitic modes) seem to be necessary to fully
address whether the main parameter determining the saturation of the
MRI is the Elsasser number or the magnetic Prandtl number.

\acknowledgments{I am grateful to Jeremy Goodman for his encouragement
  and many insightful conversations. I thank Chi-kwan Chan, Peter
  Goldreich, Yoram Lithwick, Aldo Serenelli, and Jihad Touma for
  useful discussion throughout this study.  I thank the anonymous
  referee for carefully reading the original manuscript and providing
  comments that helped improve this paper. The calculations presented
  in this paper were carried out in the Aurora Cluster at the
  Institute for Advanced Study.  I gratefully acknowledge the W.\ M.
  Keck Foundation and the Institute for Advanced Study for their
  generous support.}

\end{document}